%% file: main.tex
\documentclass[usenatbib]{mn2e}
\usepackage{epsfig}
\usepackage{amsmath,amssymb}
\usepackage{aas_macros}

\input{allvars}

\title[Low-metallicity cloud formation]
      {Metal-poor star formation triggered by the feedback effects from Pop III stars}

\author[G. Chiaki et al.]
{Gen Chiaki,$^{1}$\thanks{E-mail: chiaki@center.konan-u.ac.jp}
Hajime Susa$^{1}$ and
Shingo Hirano$^{2}$
\\
$^{1}$Department of Physics, Konan University,
8-9-1 Okamoto, Kobe, 658-0072, Japan \\
$^{2}$Department of Astronomy, The University of Texas at Austin, Austin, TX 78712, USA
}

\begin{document}

\date{}

\pagerange{\pageref{firstpage}--\pageref{lastpage}} \pubyear{2015}

\maketitle

\label{firstpage}

\begin{abstract}
Metal enrichment by the first-generation (Pop III) stars is the very first step 
of the matter cycle in the structure formation and it is followed by the formation 
of extremely metal-poor (EMP) stars. To investigate the enrichment process by the 
Pop III stars, we carry out a series of numerical simulations including the feedback 
effects of photoionization and supernovae (SNe) of Pop III stars with a range of 
masses of minihaloes (MHs), $\Mhalo$, and Pop III stars, $\MPopIII$. We find that the 
metal-rich ejecta reaches neighbouring haloes and external enrichment (EE) occurs 
when the halo binding energy is sufficiently below the SN explosion energy, $\Esn$. 
The neighbouring haloes are only superficially enriched, and the metallicity of 
the clouds is
${\rm [Fe/H]} < -5$.
Otherwise, the SN ejecta falls back and recollapses to form 
enriched cloud, i.e. internal enrichment (IE) process takes place. 
In case that a Pop III star explodes as a core-collapse SNe (CCSNe),  
MHs undergo IE, and the metallicity in the recollapsing region is $-5 \lesssim {\rm [Fe/H]} \lesssim -3$
in most cases.
We conclude that IE from a single CCSN can explain the formation of EMP stars. 
For pair-instability SNe (PISNe), EE takes place for all relevant mass range of MHs,
consistent with no observational sign of PISNe among EMP stars.
\end{abstract}


\begin{figure}
\includegraphics[width=8cm]{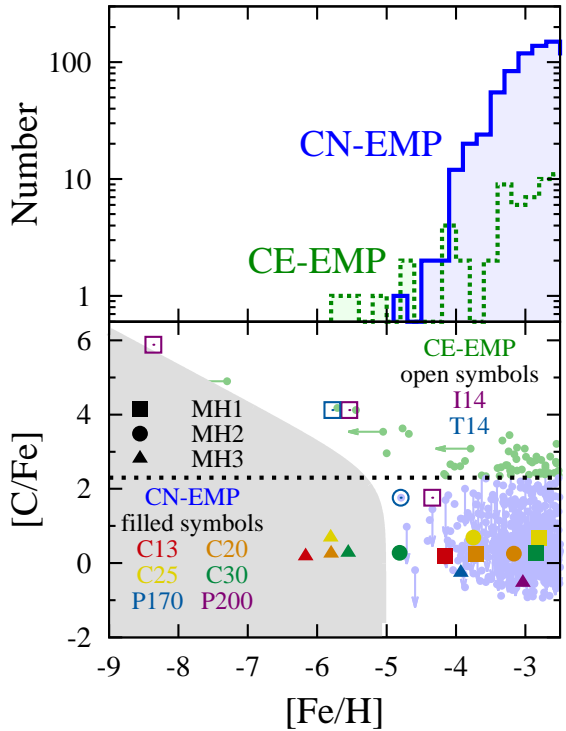}
\caption{
{\bf Top}: Number distribution of CN-EMP (blue solid curve) and 
CE-EMP (green dotted) stars as a function of metallicity [Fe/H].
{\bf Bottom}: Elemental abundance ratios of observed CN-EMP (blue dots) and CE-EMP (green)
on the [C/Fe]-[Fe/H] plane retrieved by SAGA database \citep{Suda08, Suda17}.
The horizontal black dotted line is the boundary between the two classes 
(${\rm [C/Fe]}_{\rm b} = 2.30$), and gray shaded area is the region where
the dust cooling efficiency is too small to trigger the low-mass star formation
\citep{Chiaki17}.
The symbols denote our simulation results for MH1 (squares),
MH2 (circles), and MH3 (triangles).
The results in the progenitor models reproducing elemental abundances of CN-EMP and 
CE-EMP stars are denoted by filled and open symbols with the corresponding colours
indicated by bottom-left and top-right corners of the panel, respectively (see text).
}
\label{fig:C-Fe}
\end{figure}


\section{INTRODUCTION}

The stellar initial mass function (IMF) and the evolution of elemental abundances
within $\sim 1$ billion years after the Big Bang 
have been intensively investigated to uncover the nature of
the early structure formation.
The IMF and the metal content evolve, affecting each other.
The first-generation stars (Pop III stars) born in pristine clouds 
are majorly massive ($\sim 100$--$1000 \ \Msun$)
by the lack of cooling agents other than hydrogen molecules
\citep{Bromm99, Abel02, Yoshida03, Susa14, Hirano14, Hirano15}.
They explode to enrich the surrounding gas with metals and dusts. 
The metal- and dust-enriched clouds undergo fragmentation by additional radiative cooling,
and are likely to host low-mass stars \citep[Pop II/I stars][]{Omukai00,
Schneider03, Bromm03, Frebel05}.

Those star formation and metal enrichment proceed in the early stage of the structure formation.
The standard $\Lambda$ cold dark matter ($\Lambda$CDM) model predicts that 
the structures hierarchically evolve from small to large ones, i.e., from stars to 
galaxies/galaxy clusters \citep{Audouze95}.
Pop III stars are formed in tiny dark matter (DM) haloes with mass $\Mhalo \sim 10^6 \ \Msun$ and with
the virial temperature $T_{\rm vir} \simeq 1000$ K, so-called minihaloes (MHs), at redshifts $z \sim 20$.
Then, through the halo merger, the first galaxies, defined as objects in which the cycle of material
stably stands, are formed in atomic-cooling haloes with $\Mhalo \sim 10^8 \ \Msun$
at $z\sim 10$ \citep{Bromm11}.
These objects are considered to contain more primitive population of stars and
elemental abundances than those in the present-day massive galaxies.
In the next decades, observational instruments such as the Thirty-Meter Telescopes (TMT),
Space Infrared Telescope for Cosmology and Astrophysics ({\it SPICA}), and James-Webb Space Telescope 
({\it JWST}) target the high-redshift galaxies, or even
the first galaxies.
The formation of galaxies at high redshifts have intensely been investigated by numerical simulations 
\citep[e.g.][]{Wise12, Xu13, Barrow17a, Barrow17b}
and semi-analytical studies
\citep[e.g.][]{deBennassuti14, deBennassuti17, Komiya15, Magg17}.
The confrontation of these simulations with the future observations will help us to understand the metal enrichment process in the early universe.

Another way to approach the stellar IMF and metal enrichment process in 
the early Universe is to observe long-lived metal-poor (MP) stars
residing in the outskirts of our Galaxy and ultra-faint dwarf (UFD) galaxies. 
The metallicity of MP stars are often measured as iron abundance
normalized by the solar one, 
$\abH{Fe} = \log y(\Fe) - \log y_{\bigodot}(\Fe)$,
where $y(j)$ is the number abundance of element $j$ relative to hydrogen nuclei.
In particular, the stars with $\abH{Fe} < -3$
are called as extremely metal-poor (EMP) stars, which are
considered to be enriched by a single or several SN(e) 
for their small metal contents \citep{Ryan96, Cayrel04}.
As Figure \ref{fig:C-Fe} shows, they are classified according to their carbon-enhancement 
as C-enhanced EMP (CE-EMP) and C-normal EMP
(CN-EMP) stars above and below $\abFe{C}_{\rm b} = 2.30$, respectively 
\citep{Beers05, Aoki07, Chiaki17}.
It is considered that MP stars are originated from the fragmentation of their parent 
clouds by the radiative cooling of dust thermal emission so that the low-mass, i.e., long-lived
stars are likely to be formed
\citep{Omukai00, Schneider03, Schneider06, Bromm14}.
The distribution of CE-EMP and CN-EMP stars shows the vacant region on [C/Fe]-[Fe/H] plane 
(gray shaded region), which would indicate the minimum elemental abundances 
${\rm [Fe/H]}_{\rm cr} \sim -5$ and ${\rm [C/H]}_{\rm cr} \sim -3$ 
required to activate the carbon and silicate dust cooling, respectively \citep{Chiaki17}.

The enrichment of the gas that will collapse to form EMP stars proceeds in qualitatively distinct ways.
It is considered that there are two major processes, internal enrichment (IE) and external 
enrichment (EE).
In the former case, SN shocks rich with metals fall back into and pollute the MH which has hosted a 
progenitor Pop III star. 
\citet{Ritter12, Ritter15, Ritter16} and \citet{Sluder16} 
find that, several ten million years after the explosion, the gas contracts again by the gravitational
potential of the central MH, 
and the resulting metallicity is above $10^{-4} \ \Zsun$.
In the latter case, SN shocks from a MH reach and pollute neighbouring haloes.
\citet{Smith15} report that a neighbouring halo is enriched up to 
$2\E{-5} \ \Zsun$ in a cosmological simulation.
Besides, with the initial condition put by hand and higher resolution simulations, \citet{Chen16} 
report that metal can not penetrate into the center of the clouds eroded by SN shocks
for any models but for a model with a PISN.
These simulations predict the lower metallicity in the second-generation star-forming regions
than IE.
We expect that the metallciity range depends on the enrichment processes.

Both IE and EE should occur, 
depending on the balance between the injected energies of ultraviolet photons/SN explosions  and  the halo binding energy.
Therefore, to know the conditions for the two enrichment processes and which process is dominant,
it is required to perform simulations in a wide range of parameters such as
masses of MH $\Mhalo$ and a Pop III star $\MPopIII$.
However, IE and EE have so far been separately investigated with one combination 
of the initial parameters by each earlier work.

In addition, the elemental abundance of EMP stars is the clue to restrict IMF of Pop III stars.
Pop III stars with masses $8 \ \Msun < \MPopIII < 40 \ \Msun$ explode as normal Type-II core-collapse
SNe (CCSNe) and eject the normal elemental abundance ratio.
More massive ones with $140 \ \Msun < \MPopIII < 260 \ \Msun$ explode as pair-instability SNe (PISNe)
and their ejecta show Fe-enhanced features and the large abundance contrasts between odd and even elements 
\citep{Aoki14}.
The elemental abundances of EMP stars can well be reproduced by the nucleosynthetic
models of CCSNe.
Interestingly, the stars with elemental abundances consistent with PISN model have not been
observed so far.
This is considered to be either because the formation rate of extremely massive Pop III stars are intrinsically small,
or because the ejecta from PISNe, if exist, can not enrich their host or neighbouring MHs.
Again, most of earlier works employ one initial parameter of $\MPopIII$, and 
restrict their interest to either CCSNe or PISNe.
To compare with the observations, the simulations with $\MPopIII$ covering the 
mass ranges from CCSNe to PISNe is required.

Another clue of the nucleosynthetic feature of Pop III stars is that
CE-EMP stars account for the dominant fraction (80\%) for the subset of EMP stars 
with lowest metallicity ($\abH{Fe} < -4.5$).
Their origin is considered to be the enrichment of their parent clouds by faint SNe (FSNe)
whose inner layer of ejecta undergoes mixing/fall-back and only C-rich materials are ejected
\citep{Umeda03, Marassi14, Marassi15, Chen17, Chiaki17}.
Although the model can well reproduce the relative elemental abundance of CE-EMP stars,
it has not yet been explicitly shown that the absolute elemental abundance of enriched clouds
becomes as high as observed values of $\abH{C} \sim -3$ regardless of the loss of the ejecta mass.

We in this paper focus on the first step of the matter cycle from the formation of Pop III stars to 
star-forming clouds of Pop II stars.
We perform a series of simulations to study the metal enrichment of star-forming clouds 
by a single Pop III star, considering its radiative/mechanical feedbacks.
The method is described in Section \ref{sec:method}.
We investigate two environmental parameters: masses of MHs with a range of $\Mhalo = 3\E{5}$--$3\E{6} \ \Msun$ 
and Pop III stars $\MPopIII = 13$--$200 \ \Msun$, covering the mass range of CCSNe and PISNe.
The results of the simulations separately for the combination of massive/low-mass MHs and 
CCSNe/PISNe are shown in Section \ref{sec:results}.
We find the different `modes' appear in the enrichment processes as discussed in Section 
\ref{sec:boundary}, and apply our result to the comparison with observations of EMP stars 
(Section \ref{sec:observation}).
We discuss other issues in Section \ref{sec:discussion}, and conclude the paper in Section 
\ref{sec:conclusion}.


\section{Numerical models}
\label{sec:method}

\subsection{Flow of simulations}

This work follows the process of radiative feedback from main-sequence (MS) Pop III stars
and kinetic feedback and metal dispersion from their SNe.
Throughout the simulations, we employ the $N$-body/Smoothed Particle Hydrodynamics (SPH) code \Gadget-2 \citep{Springel05}
including all relevant non-equilibrium chemistry and heating/cooling processes.
The whole course of the simulations is divided into the following three steps.\\

\noindent{\bf Formation of minihaloes and Pop III stars} 

We extract MHs from a parent cosmological simulation with a large simulation box, and 
calculate gravitational collapse of Pop III star-forming clouds hosted by MHs until 
the central density reaches $n_{\rm H,cen} = 10^3 \ \percc$ and run-away collapse onsets.\\

\noindent{\bf Radiative feedback} 

After the collapse simulations, we put a collisionless particle as a Pop III star
at the centroid of SPH particles with densities $\nH > n_{\rm H,cen} / 3$.
The velocity of the star particle is initially mass-weighted average of the central SPH particles.
We update it by integrating the gravity force.
The star particle is the source of ionization and dissociation photons.
According to their mass, the emission rates of ionizing and dissociating photons are given.
After the corresponding lifetime of a Pop III star, the photon emission is turned off
to give the initial condition of the subsequent simulations
of SN explosions.\\

\noindent{\bf Supernova feedback} 

Then, propagation of shock front associated with the supernova explosion 
is followed under the point explosion approximation.
All of the energies are added as thermal energy at the location of the Pop III star particle.
The energy is given according to the different types of SN, CCSNe and PISNe.
We follow the metal dispersion until the density of the recollapsing region reaches
$n_{\rm H,cen} = 10^3 \ \percc$ again.

\subsection{Resolution}

In an SPH scheme, a density distribution is expressed as the spatial distribution of SPH particles.
As the nature of the scheme, the spatial resolution is automatically improved with increasing density.
The employed mass of an SPH particle is $m_{\rm p}^{\rm gas} = 4.45 \ \Msun$ and the number of 
neighbouring particles is $\Nngb = 64$. As a result, the minimum mass which SPH particles resolve is $\Mmin = \Nngb \mpart ^{\rm gas}$.
For the maximum density $n_{\rm H, max}$,
the minimum length which can be resolved is described as the minimum smoothing kernel size,
\begin{equation}
h_{\rm min}
= 1.3 \ {\rm pc}
\left( \frac{\Nngb }{64} \right) ^{1/3} 
\left( \frac{\mpart ^{\rm gas}}{4.45 \ \Msun} \right) ^{1/3}
\left( \frac{n _{\rm H,max}}{10^3 \ \percc} \right)^{-1/3}.
\label{eq:resolution}
\end{equation}
However, when the size fails to sufficiently resolve the local Jeans length,
it is reported that spurious fragmentation occurs \citep{Truelove97}.
At the maximum density, the ratio $\Rcr$ of the local Jeans mass $\Mjeans$ to
$\Mmin $ is
\begin{eqnarray}
\Rcr &=& \frac{ \Mjeans}{ \Mmin } 
= 18
\left( \frac{\Nngb }{64} \right) ^{-1}
\left( \frac{\mpart ^{\rm gas}}{4.45 \ \Msun} \right) ^{-1} \nonumber \\ && \times
\left( \frac{\nH ^{\rm max}}{10^3 \ \percc} \right)^{-1/2}
\left( \frac{T}{300 \ {\rm K}} \right)^{2/3}
\end{eqnarray}
\citep{ChiakiYoshida15}.
In simulations of isolated collapsing gas clouds,
the ratio $\Rcr$ is obliged to be above 1000 by zoom-in techniques such as the particle-splitting 
\citep{Hirano14, Chiaki16}.
In our simulations, the minimum resolution of $\Rcr = 18$ is sufficient because
we are not interested in the cloud fragmentation but the average metallicity of recollapsing region.

\subsection{Chemical reactions and heating/cooling}
\label{sec:chemcool}

\subsubsection{Non-radiative reactions}
Throughout the simulation, we solve the non-equilibrium chemical networks of 53 reactions for
15 primordial species of 
$\rm e$, $\rm  H$, $\rm  H^+$, $\rm  H_2$, $\rm  H^-$, $\rm H_2^+$,
$\rm HeH^+$, $\rm  He$, $\rm  He^+$, $\rm  He^{2+}$, 
$\rm D$, $\rm  D^+$, $\rm  D^-$, $\rm  HD$, and $\rm  HD^+$,  
which cover the wide range of density of $10^{-3} \ \percc < \nH < 10^{5} \ \percc$
and temperature of $100 \ {\rm K} < T < 10^9 \ {\rm K}$.
The collisional ionization and recombination of H and He atoms and ions,
charge transfer of relevant species, and formation of H$_2$ and HD molecules are included.
The compilation of reaction rates is taken from the primordial part of \citet{Nagakura09}.

\subsubsection{Radiative cooling}
From abundances of the species, we calculate the radiative cooling rates of
the inverse Compton cooling, bremsstrahlung, 
H, He, and He$^+$ transition lines including two-photon decay,
ionization and recombination cooling,
and ro-vibrational transition line cooling of H$_2$ and HD molecules.
We do not include the radiative cooling by ejected metal because
it activates at densities $\nH \gtrsim 10^3 \ \percc$ with metallicities 
$Z \gtrsim 10^{-3} \ \Zsun$, above our maximum density \citep{Jappsen07}.

We ignore the star formation in neighbouring haloes, 
focusing on the enrichment of one halo which have hosted the Pop III star.
In order to prevent the gas from collapsing in the neighbouring haloes, 
H$_2$ and HD cooling is switched off in the region 100 pc outside from the target 
halo, the typical virial radius.
Inside 100 pc from the center of the target halo,
we prevent density from increasing larger than $n_{\rm H,cen} = 10^3 \ \percc$
by switching off molecular cooling due to the limit of resolution.

\subsubsection{Photo-chemistry and heating}

\noindent{\bf Radiative transfer}

During MS of Pop III stars, we perform radiation transfer simulations, 
using the ray-tracing method presented by \citet{Susa06}.
The rates of photoionization and dissociation are calculated at the position of each SPH particle.
Since they depend on the column density of the relevant species,
we derive this by integrating the element $\Delta N_i$(H {\sc i}) along the ray from the source to the target SPH particle.
The element $\Delta N_i$(H {\sc i}) is calculated between a SPH particle $i$ on the ray
and one of its neighbors $j$ that is selected
so that the angle between the direction from $i$ to $j$ and the direction from $i$ to the source is smallest.
The subroutine of radiative transfer are heavily parallelized.\\

\noindent{\bf Photoionization of H atoms}

We add the reactions of photoionization of H atoms and photodissociation of H$_2$ molecules.
The former proceeds by the ionizing photons with energies 
$h \nu > h \nu _{\rm L} = 13.6 \ {\rm eV}$ as
\[
\rm H + \gamma \to H^+ + e^{-}.
\]
The reaction rate $k_{{\rm ion}, i}$ and corresponding heating
rate $\Gamma _{{\rm ion}, i}$ at the position of an SPH particle $i$
are calculated from the column density $N_i$(H {\sc i}), using the photon-conserving 
procedure \citep{Susa06}.
The frequency-dependent cross-section $\sigma _{\nu}$ is taken from \citet{Shapiro87}.
We ignore the photoionization of helium atoms and ions.\\

\noindent{\bf Photodissociation of H$_2$ molecules}

The dissociation of hydrogen molecules induced by photons in the Lyman-Werner band ($11.2$--$13.6$ eV)
are also solved as
\[
\rm H_2 + \gamma \to H + H.
\]
Using the self-shielding approximation of \citet{Draine96}, we calculate the 
photodissociation rate by the column density of H$_2$ molecules.

\subsection{Supernova Feedback}

\subsubsection{Mechanical feedback}
The explosion energy is added as thermal energy uniformly in a spherical region around a Pop III remnant
which contains $N_{\rm FE} = 200$ gas particles, mimicking the end of the free-expansion (FE) phase.
Note that the FE phase is transmitted into the Sedov-Taylor (ST) phase 
when a SN ejecta sweeps up the ambient medium with the same mass.
Since the mass of SPH particles is $m_{\rm p}^{\rm gas} \sim 5 \ \Msun$ in this simulation,
the mass of the initial heated region is $N_{\rm FE} m_{\rm p}^{\rm gas} \sim 1000 \ \Msun$, which is larger than 
the ejecta mass.
It is the trade-off for the large simulation box size of $\sim \ {\rm kpc}$ to cover the entire volume
which PISN shell can cover.
Zoom-in techniques such as particle splitting \citep{ChiakiYoshida15} could resolve the innermost region of
SN ejecta.
However, when refined particles are brown away into the ambient region with coarse 
gas particles, significant error in density and pressure force estimations would be induced.
Thus, we do not employ the particle splitting in the present simulations.

\begin{table}
\begin{minipage}{8.5cm}
\caption{Properties of minihaloes}
\label{tab:MHs}
\begin{tabular}{cccccc}
\hline
         & MH  & $\zcol$ & $\Rhalo ^{\rm col}$ & $\Mhalo ^{\rm col}$ & $M_{\rm pri}^{\rm col}$ \\
         &     &         & [pc]                & [$\Msun$]           & [$\Msun$]               \\
\hline \hline                                                                                  
{\tt LH} & MH1 & 28.47   &  70.12              & $2.94\E{5}$         & $3.91\E{4}$             \\
         & MH2 & 27.52   &  79.52              & $3.89\E{5}$         & $5.26\E{4}$             \\
{\tt HH} & MH3 & 23.58   & 186.85              & $3.23\E{6}$         & $4.80\E{5}$             \\
\hline
\end{tabular}
\medskip \\
Note --- $\Mhalo ^{\rm col}$ and $\Rhalo ^{\rm col}$ denote
the virial mass and radius of MHs so that the average density of both gas and
dark matter is 200 times the critical density at the redshift $\zcol$ when 
the maximum gas density reaches $\nHcen ^{\rm col} = 10^3 \ \percc$.
$M_{\rm pri}^{\rm col}$ is the gas mass within $\Rhalo ^{\rm col}$.
\end{minipage}
\end{table}

\begin{table*}
\begin{minipage}{12cm}
\caption{Properties of Pop III stars}
\label{tab:PopIIIstars}
\begin{tabular}{ccccccccc}
\hline
           & ID         &  $M_{\rm PopIII}$ & $^{a}t_{\rm life}$ & $^{a}T_{\rm eff}$ & $^{a}Q({\rm H})$ & $^{b}\Esn$      & $^{b}\Mmet ^{\rm ej}$ & $^{b}M_{\rm Fe}^{\rm ej}$ \\
           &            &  $[\Msun]$        & [Myr]              & [K]               & [s$^{-1}$]       & [10$^{51}$ erg] & [$\Msun$]             & [$\Msun$] \\    
\hline \hline
{\tt CCSN} & {\tt C13}  &  13               & 13.7               & $5.22\E{4}$       & $1.33\E{48}$     &  1              & 0.746                 & 0.0718 \\
           & {\tt C20}  &  20               & 8.43               & $6.41\E{4}$       & $4.72\E{48}$     &  1              & 2.56                  & 0.0789 \\
           & {\tt C25}  &  25               & 6.46               & $7.08\E{4}$       & $7.58\E{48}$     &  1              & 3.83                  & 0.0758 \\
           & {\tt C30}  &  30               & 5.59               & $7.37\E{4}$       & $1.33\E{49}$     &  1              & 7.18                  & 0.105  \\
{\tt PISN} & {\tt P170} & 170               & 2.32               & $9.83\E{4}$       & $2.16\E{50}$     & 20              & 83.4                  & 4.42   \\
           & {\tt P200} & 200               & 2.20               & $9.98\E{4}$       & $2.62\E{50}$     & 28              & 114                   & 7.74\\
\hline
\end{tabular}
\medskip \\
References --- $a$. The quantities are linearly interpolated from table 4 of 
\citet{Schaerer02}; 
$b$. \citet{Umeda02}
\end{minipage}
\end{table*}

\subsubsection{Metal enrichment}

The propagation of metals is traced by Lagrangian test particles with mass 
$\mpart ^{\rm met} = \Mmet ^{\rm ej} / \Nmet$, where
$\Mmet ^{\rm ej}$ is the metal mass produced by SNe and $\Nmet$ is the number of metal particles.
We set $\Nmet = 10^5$ as in \citet{Ritter12}.
The velocity $\vb{v}$ of a metal particle $i$ is interpolated from neighbouring SPH particles $j$
with the ordinal SPH procedure as
\begin{equation}
\vb{v}_{i} = \sum _{j} \frac{m_j}{\rho _j} \vb{v}_{j} W(r_{ij}, h_{i}),
\end{equation}
where $m_j$, $\rho _j$, and $\vb{v}_{j}$ denote the mass, density, and velocity of SPH particle $j$,
$r_{ij}$ is the distance between the metal particle $i$ and SPH particle $j$, 
$h_{i}$ is the smoothing length of metal particle $i$ within which $\Nngb$ SPH particles are contained,
and $W(r,h)$ is the smoothing kernel function.
Then the position $\vb{x}_i$ of a metal particle $i$ at the time $t$ are updated as 
$\vb{x}_{i} (t+\Delta t) = \vb{x}_{i} (t) + \vb{v}_i (t)\Delta t$,
where $\Delta t$ is the simulation timestep.

Note that we here employ the ordinal cubic spline kernel to estimate metal density of each SPH particle. 
In order to obtain a better estimation of metallicity,
some researchers adopt various kernel functions \citep{Tornatore07a, Tornatore07b, Dehnen12}.
\citet{Tornatore07b} compare the runs with the ordinal cubic
spline and top-hat kernel function.
They conclude that the difference of metallicity between two runs are less significant relative to the diversity of realizations.
Thus we employ the ordinary cubic spline kernel function.

\subsubsection{Metallicity in recollapsing region}

The metallicity of the recollapsing region is calculated as
\begin{equation}
\Zcen ^{\rm recol} = \frac{M_{\rm met,cen} ^{\rm recol}}{M_{\rm pri,cen} ^{\rm recol}+ M_{\rm met,cen} ^{\rm recol}},
\end{equation}
where $M_{\rm met,cen} ^{\rm recol}$ and $M_{\rm pri,cen} ^{\rm recol}$ are the mass of metal and primordial gas particles within the local Jeans radius centered at the SPH particle of highest density.
From the total ejected mass $M_{\rm Fe}^{\rm ej}$ of Fe, we can estimate the number abundance $y _{\rm cen}^{\rm recol} ({\rm Fe})$ of Fe 
relative to hydrogen in the recollapsing region as
\begin{equation}
y _{\rm cen}^{\rm recol} ({\rm Fe}) =
\frac{Z_{\rm cen}^{\rm recol} M_{\rm Fe}^{\rm ej}}{\mu _{\rm Fe} \XH M_{\rm met}^{\rm ej}},
\label{fig:Z2Fe}
\end{equation}
where $\mu _{\rm Fe} = 56$ is the molecular weight of Fe.
Using the solar abundance $A_{\bigodot} ({\rm Fe}) = 12 + \log [y_{\bigodot}({\rm Fe})] = 7.50$
\citep{Asplund09}, the relative abundance is estimated as
$\abH{Fe}_{\rm cen}^{\rm recol} = \log [ y _{\rm cen}^{\rm recol} ({\rm Fe}) / y_{\bigodot} ({\rm Fe}) ]$.

\begin{figure*}
\includegraphics[width=15cm]{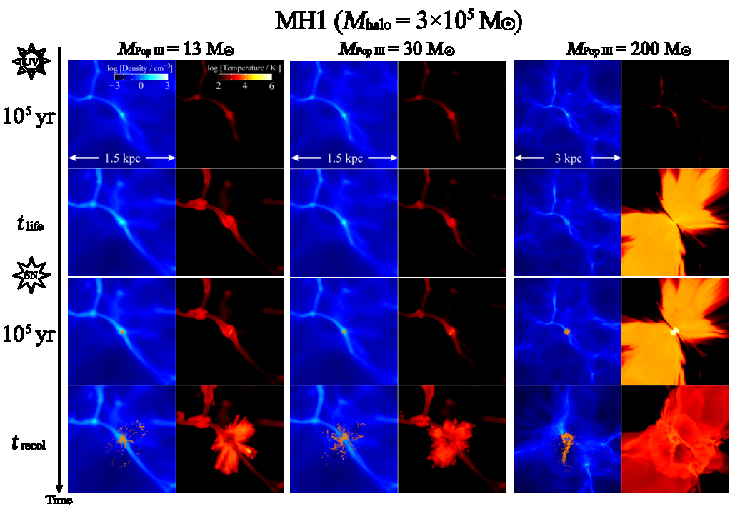}
\caption{
Density (left panels) and temperature (right) slices for MH1 at
$10^5$ yr after star formation, 
the life time of Pop III stars $t_{\rm life}$, 
$10^5$ yr after SN explosion, and 
the time when gas recollapses $t_{\rm recol}$ (from top to bottom rows).
The physical box size is indicated in the first row.
The orange dots overplotted on the density maps represent 
the position of metal particles close to the slicing plane within 2.5 times their kernel size.
}
\label{fig:snapshots_MH001}
%
\includegraphics[width=15cm]{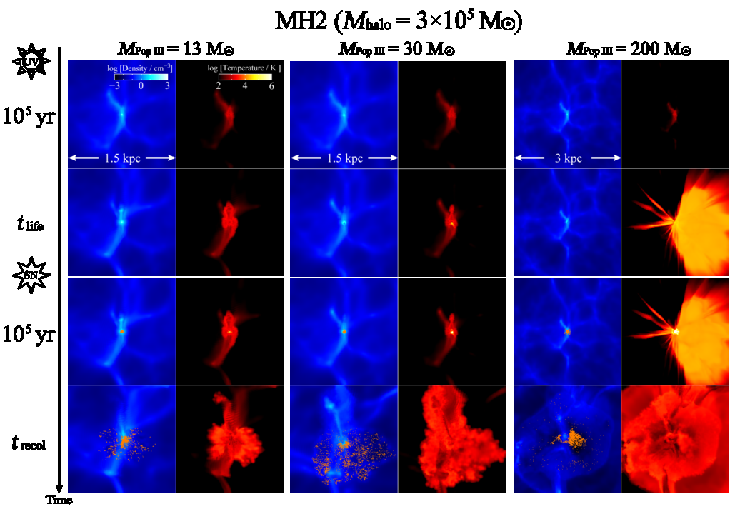}
\caption{
Same as Figure \ref{fig:snapshots_MH001} but for MH2.
}
\label{fig:snapshots_MH004}
\end{figure*}

\begin{figure*}
\includegraphics[width=15cm]{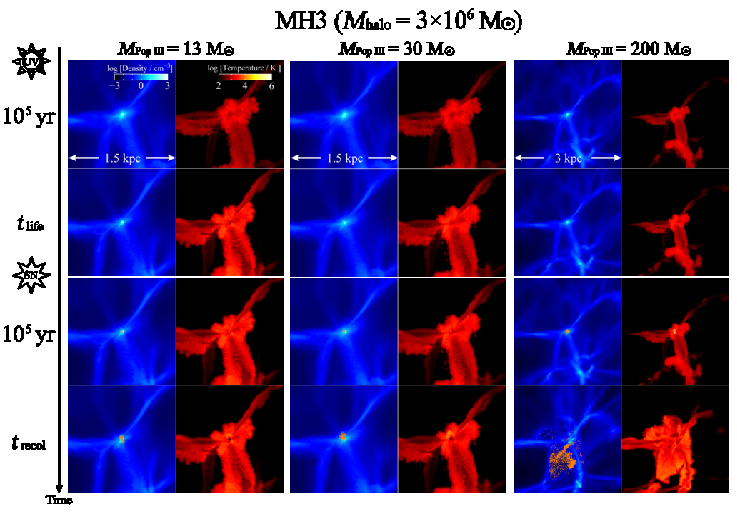}
\caption{
Same as Figure \ref{fig:snapshots_MH001} but for MH3.
}
\label{fig:snapshots_MH008}
\end{figure*}

\subsection{Initial conditions}

\subsubsection{Minihaloes}

In the parent cosmological simulation, we use \Gadget-2 suitably modified to follow the formation of primordial gas clouds \citep{Hirano14}.
The simulations are initialized at $z_{\rm ini} = 99$ with a box size $10\,h^{-1}$\,Mpc using the {\sc music} software \citep{Hahn11}.
Cosmological parameters are adopted from the latest Planck data \citep[][last column of their table 4]{Planck2015}.
We first run a dark matter only simulation with $512^3$ $N$-body particles, and identify dark haloes at $z = 10$ by running a friends-of-friends halo finder.
Next, we perform multi-level zoom-in simulations for the selected target haloes with a high spatial resolution.
Finally, we perform the zoom-in simulations including SPH particles with mass $m_{\rm p}^{\rm gas} \sim 0.3 \ \Msun$.
The SPH simulations are stopped when the target haloes collapse gravitationally, and the central gas density reaches $n_{\rm H} = 10^3\,{\rm cm^{-3}}$.
We select three MHs with masses $3\E{5}$--$3\E{6} \ \Msun$ to cover the mass range of MHs in the entire cosmological MHs given by \citet{Hirano15}.
We step back the simulation when the central gas density is $\nH = 1 \ \percc$,
and cut out a spherical region with a radius 2.5 kpc centered at the density peak so that the SN shocks even for PISNe can not go beyond the simulation region.
We then restart the collapse simulation with the chemical reactions and radiative cooling presented in Section \ref{sec:chemcool}.

Table \ref{tab:MHs} shows the total mass $\Mhalo ^{\rm col}$ and 
baryon mass $M_{\rm pri} ^{\rm col}$ within the virial radius $\Rhalo ^{\rm col}$ 
of the MHs at the time when their maximum density reaches $10^3 \ \percc$ 
for the first time.
MH1 and MH2 host Pop III stars at high redshifts $27$--$28$.
The mass of both MH1 and MH2 is around $3\E{5} \ \Msun$, 
but MH1 does not have substructures within 500 pc around it while MH2 does.
Until the time when shocked gas falls back into MH2, it is expected that 
the neighbouring haloes also merge with MH2 and reduce its metallicity.
We also study the enrichment of more massive halo, MH3, with $3\E{6} \ \Msun$ growing at a lower 
redshift 24.

\subsubsection{Pop III stars}

The mass function of Pop III stars is still under the debate 
because of the limitations of spatial resolution and computational time to follow
the entire accretion history of a disk into the central protostellar core. 
Several researchers have reported that 
the fragmentation of accretion disk around a protostar drives 
the formation of Pop III star cluster including low-mass ($\lesssim 1 \ \Msun$) stars 
\citep{Clark11, Greif11, Stacy12, Susa13, Susa14, Chiaki16, Hirano17}.
However, even in such simulations, the primary protostar is still massive ($\gtrsim 10M_\odot$). 
In the present simulations we only take into account the feedback by a primary star.

We consider the wide range of progenitor mass of 
$\MPopIII = 13$--$30 \ \Msun$ and $170$--$200 \ \Msun$ exploding as CCSNe and PISNe.
These models reproduce the elemental abundance ratio of CN-EMP stars.
The stellar properties,
such as their lifetime $\tlife$ and emission rate of ionizing photons $\QH$,
are interpolated by the quantities presented by \citet{Schaerer02}
as listed in Table \ref{tab:PopIIIstars}.
The explosion energy $\Esn$ is fixed for CCSNe to be $1\E{51} \ \erg$,
and $20\E{51}$ and $28\E{51} \ \erg$ for PISNe with $\MPopIII = 170$ 
and $200 \ \Msun$, respectively.
With those explosion energies, the mass $\Mmet ^{\rm ej}$ of nucleosynthesized metal is 
listed in Table \ref{tab:PopIIIstars} \citep{Umeda02}.
The mass of synthesized iron $M_{\rm Fe}^{\rm ej}$ is 
less sensitive to the progenitor mass than the total metal mass.
The oxygen mass accounts for the large fraction of the ejected metal, and increases 
with $\MPopIII$, but it is observationally difficult to constrain the oxygen 
abundance of EMP stars \citep{Cayrel04}.

Hereafter, we indicate these models by the combination of IDs of MHs and $\MPopIII$.
The name `{\tt MH1-C30}' denotes the model with $\MPopIII = 30 \ \Msun$ hosted in MH1, for example.
For the later discussion, we generalize the models with
$13 \ \Msun \leq \MPopIII \leq 30 \ \Msun$ as `{\tt CCSN}' and
$170 \ \Msun \leq \MPopIII \leq 200 \ \Msun$ as `{\tt PISN}'.
Also, we call low-mass MH1 and MH2 as {\tt LH} while high-mass MH3 as {\tt HH} 
to compare their hydrodynamic evolutions.

\subsubsection{Parameters of the Cosmological Simulation}
Throughout the simulations as well as the parent cosmological simulation,
the cosmological parameters are
$\Omega _{\rm m} = 0.31$,
$\Omega _{\rm b} = 0.047$,
$\Omega _{\rm \Lambda} = 0.69$, and
$H_0 = 100 \ h = 68 \ {\rm km \ s^{-1} \ Mpc^{-1}}$ \citep{Planck2015}.
We perform simulations in the comoving coordinate, 
but we show the physical quantities in proper coordinate in the text, tables, and figures hereafter.

Although helium is nucleosynthesized and dispersed by a SN, 
we fix the mass fraction of helium to $Y_{\rm He} = 0.24$. 
We show gas density in the form of number density of hydrogen nuclei as
$\nH = \XH \rho / \mH$, where $\XH = 1-Y_{\rm He}$ is the mass fraction of hydrogen nuclei,
$\rho$ is the mass density of gas, and $\mH$ is the mass of a hydrogen nucleus.
The metallicity is shown in the unit of the solar one $\Zsun = 0.02$.

\begin{figure*}
\includegraphics[width=16cm]{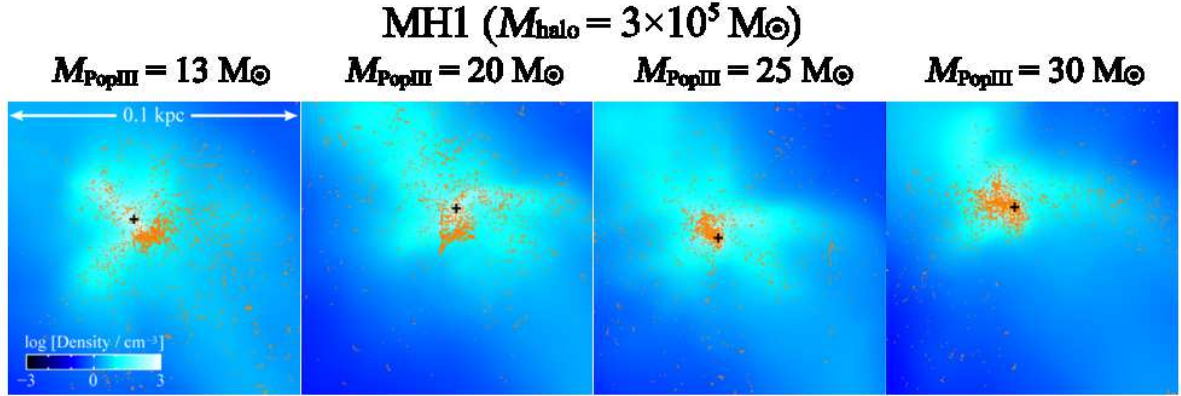}
\caption{
Central part of Figure \ref{fig:snapshots_MH001} at the time $\trecol$ after SN explosions with progenitor masses $\MPopIII = 13$--$30 \ \Msun$
for MH1 with a box size 0.1 kpc around the explosion center.
The plus symbols and orange dots indicate the positions of the gas density peak and metal particles, respectively. 
}
\label{fig:snapshots_MH001_CCSN_zoom}
\end{figure*}

\begin{table*}
\begin{minipage}{15cm}
\caption{Properties and metallicities of self-enriched haloes}
\label{tab:results}
\begin{tabular}{ccccccccc}
\hline
Halo & $M_{\rm PopIII}$ & $\nHcen ^{\rm life}$ & $\trecol$ & $\Zcen ^{\rm recol}$ & ${\rm [Fe/H]}_{\rm cen}^{\rm recol}$ & $f_{\rm halo}^{\rm met}$ & $f_{\rm halo}^{\rm pri}$ & $\Zhalo ^{\rm recol}$ \\
     & [$\Msun$]        & [$\percc$]           & [Myr]     & [$\Zsun$]            &                                      &                          &                          & [$\Zsun$]             \\
\hline \hline                                               
 MH1 &  13              & $3.14\E{2}$          & 5.74      & $4.89\E{-5}$         & $-4.16$                              & 0.555                    &  3.87                    & $1.37\E{-4}$          \\
     &  20              & $7.16\E{2}$          & 7.19      & $4.22\E{-4}$         & $-3.71$                              & 0.459                    &  2.85                    & $5.28\E{-4}$          \\
     &  25              & $2.66\E{2}$          & 8.41      & $5.31\E{-3}$         & $-2.81$                              & 0.368                    &  2.60                    & $6.94\E{-4}$          \\
     &  30              & $1.62\E{1}$          & 12.2      & $6.62\E{-3}$         & $-2.84$                              & 0.398                    &  2.62                    & $1.39\E{-3}$          \\
\hline
 MH2 &  13              & $8.31\E{1}$          & 11.3      & $4.88\E{-5}$         & $-4.16$                              & 0.701                    &  3.79                    & $1.31\E{-4}$          \\
     &  20              & $5.90\E{0}$          & 12.5      & $1.49\E{-3}$         & $-3.17$                              & 0.648                    &  2.90                    & $5.44\E{-4}$          \\
     &  25              & $6.09\E{1}$          & 22.4      & $6.03\E{-4}$         & $-3.75$                              & 0.300                    &  4.07                    & $2.68\E{-4}$          \\
     &  30              & $1.39\E{0}$          & 39.7      & $7.10\E{-5}$         & $-4.81$                              & 0.254                    &  5.96                    & $2.91\E{-4}$          \\
\hline
 MH3 &  13              & $9.02\E{5}$          & 0.979     & $4.77\E{-7}$         &($-6.17$)                             & 0.895                    &  1.91                    & $3.63\E{-5}$          \\
     &  20              & $7.88\E{5}$          & 1.16      & $3.48\E{-6}$         &($-5.80$)                             & 0.608                    &  1.31                    & $1.24\E{-4}$          \\
     &  25              & $7.31\E{5}$          & 1.23      & $5.33\E{-6}$         &($-5.80$)                             & 0.557                    &  1.32                    & $1.68\E{-4}$          \\
     &  30              & $4.77\E{5}$          & 1.30      & $1.29\E{-5}$         &($-5.55$)                             & 0.421                    &  1.30                    & $2.43\E{-4}$          \\
     & 170              & $6.30\E{2}$          & 5.85      & $1.49\E{-4}$         & $-3.93$                              & 0.368                    &  1.17                    & $2.73\E{-3}$          \\
     & 200              & $2.96\E{1}$          & 28.5      & $9.10\E{-4}$         & $-3.04$                              & 0.449                    &  2.94                    & $1.83\E{-3}$          \\
\hline
\end{tabular}
\medskip \\
Note --- 
(3) $n_{\rm H,cen}^{\rm life}$: central number density at the lifetime of a Pop III star.
(4) $\trecol$: time to recollapse from a SN explosion.
(5, 6) $Z_{\rm cen}^{\rm recol}$, ${\rm [Fe/H]}_{\rm cen}^{\rm recol}$: metallicity and iron abundance in the recollapsing cloud. Parentheses are attached for models with ${\rm [Fe/H]}_{\rm cen}^{\rm recol}$ below the critical value ${\rm [Fe/H]}_{\rm cr} = -5$ \citep{Chiaki15, Chiaki17}.
(7) $f_{\rm halo}^{\rm met}$: mass fraction of metal remaining within the virial radius $R_{\rm vir}$ to ejected.
(8) $f_{\rm halo}^{\rm pri}$: mass fraction of primordial gas within $R_{\rm vir}$ at $\trecol$ to the initial.
(9) $Z_{\rm halo}^{\rm recol}$: metallicity within $R_{\rm vir}$ at $\trecol$.
\end{minipage}
\end{table*}

\section{Results}
\label{sec:results}

\subsection{Overview}
Figures \ref{fig:snapshots_MH001}--\ref{fig:snapshots_MH008} show the density and temperature distributions during MS (top two rows)
and after SN explosions (bottom two rows) for each MH.
In most models, the central gas clouds hosted by MHs are not affected or partly broken by the emission of ionizing photons.
Then, after SN explosions, metal particles expel mainly in the direction perpendicular to the filaments of the cosmological large-scale structures, i.e., direction to the voids.
A fraction of metals is halted to be dispersed by the continuous gas accretion in the direction of the filament, and 
falls back into the central MHs at 1--10 Myr after explosions.
The MHs are self-enriched (IE) by Pop III stars which they have hosted
(Figure \ref{fig:snapshots_MH001_CCSN_zoom}).
The metallicity ranges in the recollapsing clouds are
$Z_{\rm cen}^{\rm recol} \sim 10^{-4}$--$10^{-2} \ \Zsun$ for {\tt LH}, 
consistent with that of observed EMP stars, and 
$10^{-6}$--$10^{-5} \ \Zsun$ for {\tt HH} 
as shown in Figure \ref{fig:MpopIII-Zcen_int} and Table \ref{tab:results}.
The metallicity of recollapsing regions depends not only on the halo mass and the Pop III mass, but also on the presence of substructures.
When subclumps merge before the run-away collapse of the recollapseing cloud, the pristine gas flows into the central clouds, and the metallciity becomes smaller than in the cases without substructures.

On the other hand, for {\tt LH-PISN}, H {\sc ii} regions expel to more than 1 kpc 
away from the central Pop III stars as indicated by the region with several tens-thousands kelvins.
The gas with metals is ejected from SNe and never falls back until at least 50 Myr after SN.
Instead, it goes through the voids to reach the regions within virial radii of several neighbouring haloes locating at $\sim 1$ kpc distant from the central MHs (Figures \ref{fig:snapshots_MH001} and \ref{fig:snapshots_MH004}).
We observe that metals are incorporated into the central part of one neighbouring halo, and its metallicity is $\lesssim 10^{-6} \ \Zsun$, indicating that EE is ineffective relative to IE.

The metallicity ranges in the recollapsing regions depend largely on enrichment processes, IE or EE.
We can see this for the first time by performing simulations for ranges of $\Mhalo$ and $\Esn$ ($\MPopIII$).
We see the details of hydrodynamical evolution and resulting metallicity range of the recollapsing 
regions for IE (Section \ref{sec:IE}) and EE events (Section \ref{sec:EE}).

\subsection{Internal Enrichment}
\label{sec:IE}
We observe IE for all models but for {\tt LH-PISN}.
Figure \ref{fig:rn} shows the radial density profiles (a) during MS and (b) after SN explosions 
centered at Pop III stars and their explosion center, respectively.
Radial profiles in two representative directions are shown by the solid and dotted curves 
that correspond to the directions to cosmological filaments and voids, respectively.
The dynamics of the gas in the halo and the metallicity in the recollapsing region 
depend on $\MPopIII$, $\Mhalo$, and the presence substructures.
We classify our models into three groups of {\tt LH-CCSN} (Section \ref{sec:LH-CCSN}), 
{\tt HH-CCSN} (Section \ref{sec:HH-CCSN}), and  {\tt HH-PISN} (Section \ref{sec:HH-PISN}). 
We summarize the complicated dependency of metallicity on $\Mhalo$ and $\MPopIII$ 
in Section \ref{sec:summary_dependency}.

\subsubsection{CCSNe in Low-mass MHs}
\label{sec:LH-CCSN}

In the direction of filaments, H {\sc ii} regions is still R-type at $10^6$ yr after star formation, and then 
transit to the D-type at $t_{\rm life}$ finding the D-type shocks at 1--10 pc from the Pop III stars for {\tt LH} 
(solid yellow and orange curves in Figure \ref{fig:rn}).
In the direction of voids, D-type fronts propagate more rapidly, reaching $\sim 20$ pc at $\tlife$.
The distance of the shocks from Pop III star increases with increasing $\MPopIII$ because 
the emission rates of ionizing photons increases (Table \ref{tab:PopIIIstars}).
For {\tt MH1-C30} and {\tt MH2-C30},
ionization region expands to 5 and 10 pc, and
the density at the position of the Pop III star decreases 
from the initial value of $\nHcen ^{\rm col} = 10^3 \ \percc$ to $\nHcen ^{\rm life} = 16.2$ 
and $1.39 \ \percc$, respectively.
On the other hand, for {\tt MH1-C13} and {\tt MH2-C13},
the gas density only slightly decreases to $314$ and $83.2 \ \percc$, respectively.
As the Pop III stellar mass decreases, the decrement of $\nHcen ^{\rm life}$ becomes smaller because
the ionization photon emission rate declines,
and also because the halo mass becomes larger due to accretion through the longer stellar life time.

The blast wave from SN explosion further destroy the clouds in the MHs (Figure \ref{fig:rn}b).
At $t\sim 10^6$ yr after explosion, the central region is significantly evacuated with
densities less than $< 0.01 \ \percc$.
As shown in the fourth column of Table \ref{tab:results}, the density of the recollapsing regions
reaches $10^3 \ \percc$ at the time $\trecol = 5$--$40$ Myr after the explosion.
The time $\trecol$ becomes longer with increasing $\MPopIII$ because the shock 
velocity decreases going through the denser region owing to weaker pre-ionization rates.
Although about a half of metals is dispersed in the direction of voids
(orange dots in Figure \ref{fig:snapshots_MH001}),
the rest remains around the loci of the explosion center
(Figure \ref{fig:snapshots_MH001_CCSN_zoom}).

\begin{figure}
\includegraphics[width=8cm]{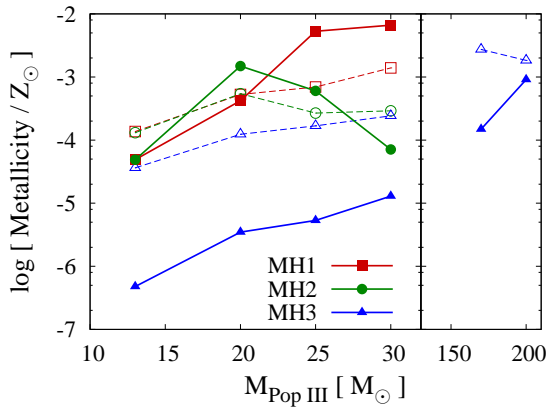}
\caption{
Metallicity in the recollapsing region
as a function of the Pop III stellar mass for MH1 (red curves), MH2 (green), and MH3 (blue). 
The solid and dashed curves show the average metallicity within the Jeans length
and virial radius, respectively.
}
\label{fig:MpopIII-Zcen_int}
\end{figure}

\begin{figure*}
\includegraphics[width=8.5cm]{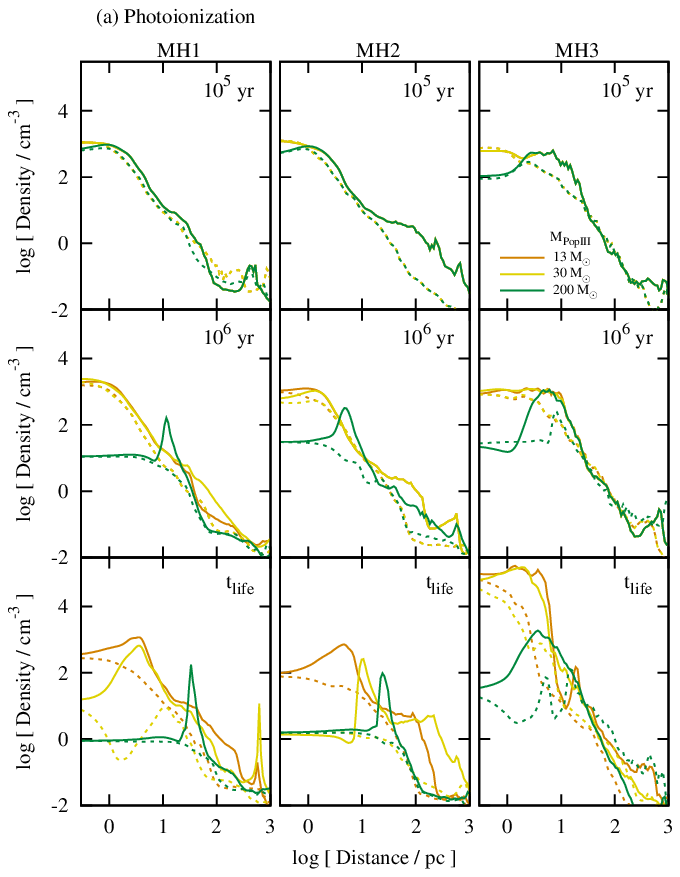}
\includegraphics[width=8.5cm]{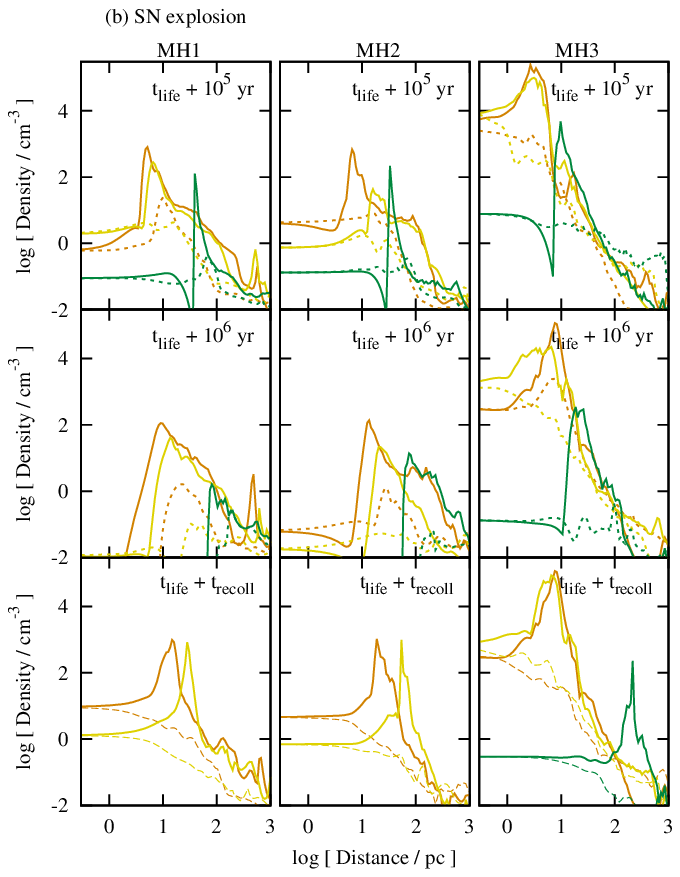}
\caption{
Gas density as a function of distance from
Pop III stars 
with masses $\MPopIII = 13 \ \Msun$ (orange curves), $20 \ \Msun$ (yellow), and
$200 \ \Msun$ (green)
hosted by minihaloes MH1 (left column), MH2 (middle), and MH3 (right).
The solid and dotted curves indicate the density profile in the directions with maximum and minimum
column densities at the radius of $\Rhalo$, respectively.
The results at the time (a) $10^5$, $10^6$ yr, and the lifetime $\tlife$ of Pop III stars during their main-sequence,
and (b) $10^5$, $10^6$ yr, and the time $\trecol$ when clouds collapse again after the SN explosions 
are shown from top to bottom row.
We will show the density profiles for $\MPopIII = 170$ and $200 \ \Msun$ in MH1 and MH2 at $t = \tlife + \trecol$ in Figure \ref{fig:rnz_ext}.
}
\label{fig:rn}
\end{figure*}

As shown in the last column in Table \ref{tab:results} and Figure \ref{fig:MpopIII-Zcen_int},
the metallicity in the recollapsing region is $5\E{-5} \ \Zsun < \Zcen ^{\rm recol} < 7\E{-3} \ \Zsun$ 
and $-4.16 < {\rm [Fe/H]}_{\rm cen}^{\rm recol} < -2.84$, which indicates that
the second generation
stars are expected to form in the enriched clouds,
to be found as EMP stars in the local Universe.
We conclude that a single CCSN is enough to pollute the halo having hosted its progenitor
to reproduce the metallicity range of the star-forming clouds of EMP stars.

The metallicity becomes smaller as the progenitor mass gets smaller.
We find the enrichment process is different between large
and small $\MPopIII$.
As shown in Figure \ref{fig:snapshots_MH001_CCSN_zoom}, for {\tt MH1-C25} and {\tt C30}, 
the shocked gas contracts again, gathering metals remaining at the central region.
On the other hand, for {\tt C13} and {\tt C20}, the gas in the direction of the filament 
is compressed and begins to collapse too rapidly to fall back again into the center
because the gas has been dense during MS due to the week ionization efficiency.
As \citet{Chiaki13} predict by one-dimensional calculations,
SN shells become gravitationally unstable during expanding in the dense medium.
Since the metals coming from behind the shock can not efficiently penetrate into the shock-driven
recollapsing region,
the only slightly enriched cloud is formed in the offset region against the explosion center.
We call the former enrichment process as the {\it central recollapse}, and 
the latter as the shock-driven {\it off-central recollapse}.

\begin{figure}
\includegraphics[width=8.5cm]{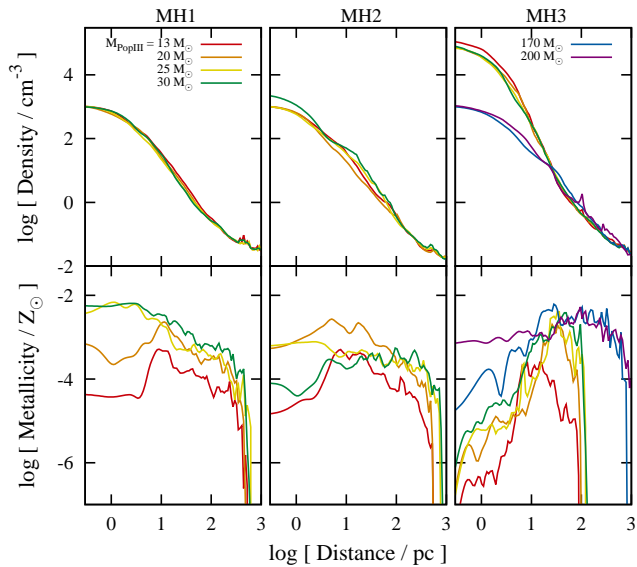}
\caption{
Gas density (top panels) and metallicity (bottom) as a function of distance from 
the collapse center for MH1 (left column), MH2 (middle), and MH3 (right) with
progenitor masses $\MPopIII = 13$--$200 \ \Msun$ from red to purple
at the recollapsing time $t = \tlife + \trecol$.
We will show the results for $\MPopIII = 170$ and $200 \ \Msun$ in MH1 and MH2 
in Figure \ref{fig:rnz_ext}.
}
\label{fig:rnz_int}
\end{figure}

Figure \ref{fig:rnz_int} shows the density and metallicity as a function of the distance from the
density maxima of the recollapsing clouds at the time $\trecol$.
For {\tt MH1-C13} (red curve), the metallicity is $\simeq 5\E{-5} \ \Zsun$ in the center
while it is higher ($\simeq 2\E{-4} \ \Zsun$) at the distance $\sim 10$ pc, which indicates that
metal does not penetrate into the off-center recollapsing region.
On the other hand, for {\tt C30} (green curve), the metallicity is higher in the center than in the ambient gas, which indicates that metal is sufficiently incorporated into the central part of the recollapsing clouds.
The slope of metallicity against the radius continuously becomes opposite with 
increasing $\MPopIII$ from red to green curves for MH1 and from red to orange 
curves for MH2 in Figure \ref{fig:rnz_int}.

We can not see this trend for {\tt MH2-C25} and {\tt C30}.
In those cases, a neighbouring halo containing only the pristine gas merges with 
the central halo from above in Figure \ref{fig:snapshots_MH004}.
For {\tt C30}, the neighbouring halo goes through 
the central one with a part of primordial gas accreted (Figure \ref{fig:snapshots_MH004}).
Consequently, the metallicity for {\tt C25} and {\tt C30} is
$6.03\E{-4}$ and $7.10\E{-5} \ \Zsun$, respectively, below the one for {\tt C20}.
The presence of substructures in the MHs can leads to the scatter 
of metallicity by two orders of magnitude.

Besides, for {\tt MH2-C13} and {\tt C20}, the neighbouring halo approaches the central 
MH, but they do not merge until the central MH recollapses because $\trecol$ is 
shorter for the less efficient photoionization than {\tt C25} and {\tt C30}.
The resulting metallicities with the same $\MPopIII$ for MH1 and MH2 are 
similar, within a factor of three.
Unless the subclumps merge, the hydrodynamic evolution
are unaffected by the substructures because they are generally connected with the central MH
through cosmological filaments by which the expansion of
the H {\sc ii} region and SN shells is halted.

Since the approaching halo is only 63.5 and 105 pc away from MH2,
we expect that EE occurs.
For {\tt C13}, the metallicity in the region centered at the density peak with 
$47.4 \ \percc$ within the Jeans radius $R_{\rm J} = 30.9$ pc is $1.32\E{-7} \ \Zsun$, 
much smaller than the critical metallicity.
For {\tt C20}, we can not observe metal particles in the neighbouring cloud with maximum 
density $214 \ \percc$ and Jeans radius $R_{\rm J} = 10.4$ pc.
\citet{Smith15} perform simulations with a neighbouring cloud at $\sim 250$ pc from a 
Pop III star-hosting cloud with $\Mhalo = 5\E{5} \ \Msun$, $\MPopIII = 40 \ \Msun$, 
and $\Esn = 1\E{51}$ erg, and report the cloud is polluted with $2\E{-5} \ \Zsun$.
This higher metallciity than we obtain may be due to the larger $\MPopIII$ with larger 
efficiency of photoionization.

We estimate the number fraction of MHs harbouring substructures.
We find $\simeq 30$ MHs which have subclumps within 250 pc among 110 cosmological samples of
\citet[][see figure D.1 of \citet{Hirano15D}]{Hirano14}.
Although further calculations are required to see whether they merge with the central MHs before
the shocked gas recollapses or not,
we can see that 30\% of MHs will evolve as MH2.

\begin{table}
\begin{minipage}{8cm}
\caption{Properties of externally enriched haloes}
\label{tab:results_ext}
\begin{tabular}{cccccc}
\hline
Halo & $M_{\rm PopIII}$ & Cloud & $D$  &  $\Mhalo ^{\rm ngb}$   & $\Rhalo ^{\rm ngb}$ \\
     & [$\Msun$]        &       & [pc] &  [$\Msun$]  & [pc] \\
\hline \hline
 MH1 & 170              & 1A    & 281  & $1.73\E{7}$ & 357 \\   
     &                  & 1B    & 909  & $2.72\E{6}$ & 192 \\   
     &                  & 1C    & 393  & $6.05\E{6}$ & 251 \\   
     &                  & 1D    & 656  & $1.77\E{6}$ & 167 \\   
     &                  & 1E    & 648  & $1.01\E{6}$ & 138 \\   
     &                  & 1F    & 903  & $1.54\E{6}$ & 159 \\   
     &                  & 1G    & 593  & $8.22\E{6}$ & 278 \\   
\hline                                 
     & 200              & 1A    & 210  & $2.14\E{7}$ & 392 \\   
     &                  & 1B    & 873  & $3.21\E{6}$ & 208 \\   
     &                  & 1C    & 335  & $1.54\E{6}$ & 351 \\   
     &                  & 1D    & 613  & $2.29\E{6}$ & 186 \\   
     &                  & 1E    & 550  & $1.07\E{6}$ & 144 \\   
     &                  & 1F    & 886  & $2.66\E{6}$ & 195 \\   
     &                  & 1H    & 606  & $1.62\E{6}$ & 165 \\   
\hline                                 
 MH2 & 170              & MH2   & ---  & $3.03\E{6}$ & 211 \\   
     &                  & 2A    & 643  & $3.14\E{6}$ & 213 \\   
     &                  & 2B    & 613  & $1.85\E{6}$ & 179 \\   
     &                  & 2C    & 1099 & $2.88\E{6}$ & 207 \\   
     &                  & 2D    & 1006 & $7.40\E{5}$ & 132 \\   
\hline                                 
     & 200              & MH2   & ---  & $2.93\E{6}$ & 210 \\   
     &                  & 2A    & 632  & $3.20\E{6}$ & 216 \\   
     &                  & 2B    & 615  & $1.86\E{6}$ & 180 \\   
     &                  & 2E    & 1312 & $7.35\E{6}$ & 285 \\   
     &                  & 2F    & 1306 & $4.25\E{5}$ & 110 \\   
\hline
\end{tabular}
\medskip \\
(4) $D$: distance of enriched neighbouring halo from the central MH.
(5, 6) $M_{\rm halo}^{\rm ngb}$, $R_{\rm halo}^{\rm ngb}$: virial mass and radius of a neighbouring halo.
\end{minipage}
\end{table}

\subsubsection{CCSNe in Massive MHs}
\label{sec:HH-CCSN}

For the massive halo (MH3; $3\E{6} \ \Msun$), the \HII \ region can not expand prior 
to the CCSNe (see Figures \ref{fig:snapshots_MH008} and \ref{fig:rn}) because the 
gas accretion is more rapid than the expansion of the H {\sc ii} region.
The central density $\nHcen ^{\rm life}$ at $\tlife$ just before the SNe grows up to
$4.77$--$9.02 \E{5} \ \percc$ (Table \ref{tab:results}) although
we turn off molecular cooling above $10^3 \ \percc$ because the gas temperature 
is still below the virial temperature $T_{\rm vir} \sim 1000$ K.

At $10^6$ yr after the SN explosions, large pressure reduces the gas density in 
the central region, but it is not sufficient to lead the complete evacuation.
While the SN ejecta barely expels in the direction of voids,
blast waves are dammed and further push the gas along the filaments.
The gas on the root of the filament collapses more rapid than the metal mixing, 
i.e.,
the off-central recollapse occurs
as we have seen in the previous section ({\tt LH-C13} and {\tt LH-C20}).
Figure \ref{fig:snapshots_MH008} shows that the distributions of metal particles 
(orange dots) remaining center with offset to the density peak.
The red to green curves in Figure \ref{fig:rnz_int} have the steeper dependency 
on radius than in {\tt LH-CCSN}.
Consequently, the metallicity in the self-enriched region becomes 
$4.77\E{-7} \ \Zsun < \Zcen^{\rm recol} < 1.29\E{-5} \ \Zsun$ and $-6.17 < {\rm [Fe/H]} < -5.55$.
This result implies that the next generation stars formed in this recollpased cloud 
would be similar to the Pop III stars because clouds with metalliciteis below the 
critical ${\rm [Fe/H]}_{\rm cr} = -5$ 
will collapse in a single or several massive object(s) due to insufficient radiative cooling
\citep{Omukai00, Schneider03, Chiaki16}.
We revisit this issue later in Section \ref{sec:critical_metallicity}.

\subsubsection{PISN in Massive MHs}
\label{sec:HH-PISN}

For the massive halo (MH3),
the \HII \ region can expand only $\sim 1$ pc, albeit the copious emission of ionization photons
($\QH \sim 10^{50} \ {\rm s^{-1}}$) for PISNe. That is caused by  the large accretion rate into the massive halo (see Figures \ref{fig:snapshots_MH008} and \ref{fig:rn}).

At $10^6$ yr after the SN explosions, large pressure leads the gas in the central region to be
evacuated as for the low-mass MHs.
For {\tt P170}, the recollapse is delayed to 5.85 Myr because of the energetic explosion.
For {\tt P200}, the gas recollapses 
at the time 28.5 Myr after the SN.

The metallicity in the recollapsing region is 
$Z_{\rm cen}^{\rm recol} = 1.49\E{-4} \ \Zsun$ and $9.10\E{-4} \ \Zsun$ 
(${\rm [Fe/H]}_{\rm cen}^{\rm recol} = -3.93$ and $-3.04$)
for {\tt P170} and {\tt P200}, respectively, above the critical metallicity ${\rm [Fe/H]}_{\rm cr}$ 
for the cloud fragmentation so that
low-mass EMP stars with the abundance ratio of PISNe 
which can survive until the present day are likely to be formed.
Following the recommendation of \citet{Karlsson08}, we estimate
the calcium abundance ${\rm [Ca/H]}_{\rm cen}^{\rm recol} = -3.26$ and $-2.25$, respectively.
We also confirm that a PISN can create the highly enriched region due to its large 
mass of ejected metals as \citet{Karlsson08} point out.
We estimate the number fraction of MHs which will host such stars
in Section \ref{sec:numfrac}.

\subsubsection{Summary of dependency of metallicity on $\Mhalo$ and $\MPopIII$}
\label{sec:summary_dependency}

\begin{figure*}
\includegraphics[width=12cm]{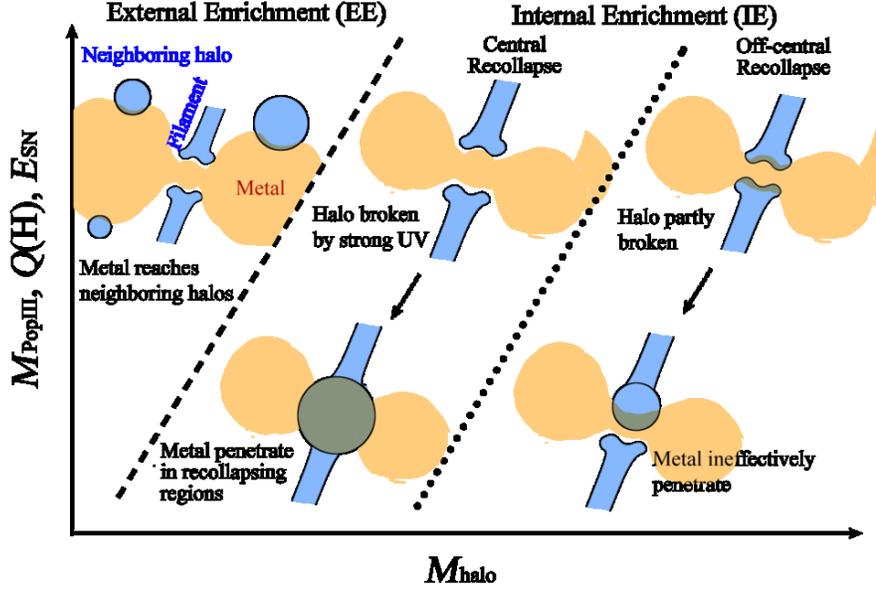}
\caption{
Schematic picture of the different modes:
enrichment processes of a cloud which collapse after the SN explosion of a Pop III star.
With increasing $\MPopIII$ and decreasing $\Mhalo$, 
metal (orange coloured region) is dispersed in the neighbouring haloes (blue circles) 
because the explosion energy overcomes the halo's binding energy (External Enrichment: EE).
Otherwise, gas falls back to the central halo having hosted the Pop III star
along the filamentary cosmic large-scale structure (Internal Enrichment: IE).
Normally, the gas recollapses to the halo center with metals (central recollapse).
For decreasing $\MPopIII$ and increasing $\Mhalo$, the region compressed by shock
in the direction of filaments collapse more rapidly than it falls back into the 
center, and acquires metals inefficiently (off-central recollapse).}
\label{fig:sche}
\end{figure*}

As we have seen so far, the metallicity range resulting from the self-enrichment 
of MHs spreads a wide range from ${\rm [Fe/H]}_{\rm cen}^{\rm recol} = -6.17$ to 
$-2.84$, depending on $\Mhalo$, $\MPopIII$ and the presence of substructures.
In case we consider large $\Mhalo$ and small $\MPopIII$, H {\sc ii} region are `extinguished' by 
the rapid gas accretion along the filaments, and by succeeding SN explosion metal 
can not sufficiently penetrate into the recollapsing region at the limb of the MH 
connecting to the filaments as schematically shown in Figure \ref{fig:sche}
(off-central recollapse).
This occurs for {\tt LH-C13} and {\tt C20}, and resulting metallicity is below 
${\rm [Fe/H]}_{\rm cen}^{\rm recol} = -3$.
Especially for {\tt HH-CCSN}, the reduction of metallicity is remarkable to be below 
${\rm [Fe/H}]_{\rm cen}^{\rm recol} = -5$.
In this metallicity range, no CN-EMP stars have been observationally found so far
(Figure \ref{fig:C-Fe}).
Therefore, we hereafter call this enrichment mode as the `ineffective IE' mode
as a subset of the cases with the off-central recollapse.
We will estimate the condition for the mode in Section \ref{sec:boundary_IE-a_IE-b}.

A series of earlier works reports that the recollapsing regions formed through IE 
have above the critical metallicity \citep{Ritter12, Ritter15, Ritter16, Sluder16}.
They however consider that $\QH = 6\E{49}$ s$^{-1}$, corresponding to 
$\MPopIII = 60 \ \Msun$ \citep{Schaerer02} and $\Mhalo ^{\rm col} = 1\E{6} \ \Msun$, 
where H {\sc ii} region sufficiently break the Pop III hosting cloud in MHs.
By performing simulations with lower $\MPopIII$ and higher $\Mhalo$, we find the 
ineffective IE mode.

\subsection{External enrichment}
\label{sec:EE}

\noindent{{\it PISN in Low-mass MHs}}\\

In the case of {\tt LH-PISN},
Figures \ref{fig:snapshots_MH001} and \ref{fig:snapshots_MH004} show that \HII \ region 
expands beyond the virial radius ($\sim 100$ pc), since the halo mass is less than the 
critical mass.
Ionizing photons expel to the void regions, but
the gas temperature does not change along the filaments of the cosmic web,
which leads the shape of \HII \ region to be bipolar.
The gas density decreases down to $\sim 1 \ \percc$ due to strong pressure as shown 
by red and green curves in Figure \ref{fig:rn}, which is consistent with the one-dimensional 
calculations of \citet{Kitayama04} and \citet{Whalen08}.

\begin{figure}
\includegraphics[width=8cm]{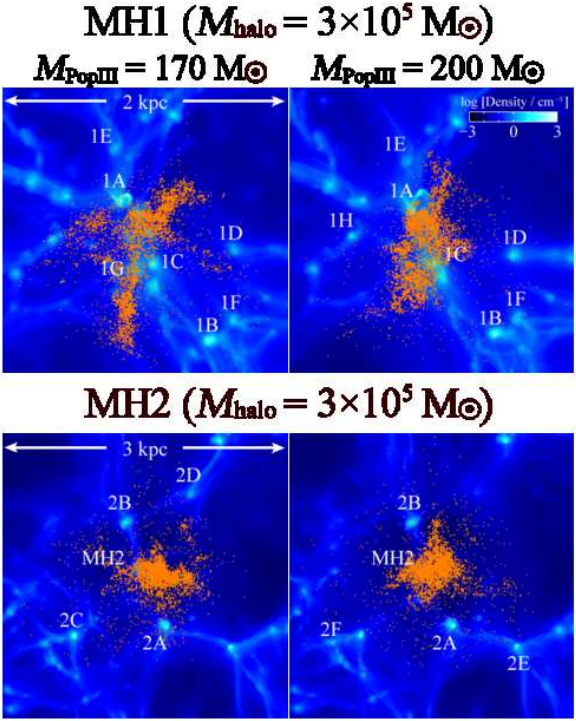}
\caption{
Density projection for PISNe with progenitor masses of $\MPopIII = 170$ and $200 \ \Msun$ 
in low-mass haloes (MH1 and MH2) at $\simeq 50$ Myr after SNe.
Labels show
the halo having hosted the Pop III star and 
neighbouring haloes within whose virial radius the average metallicity is non-zero.
}
\label{fig:snapshot_ext}
\end{figure}

Figure \ref{fig:snapshot_ext} shows the projection of density for {\tt LH-PISN}.
Since the SN shocks go through the rarefied regions, the enriched region extends
to $\sim 1$--$1.5$ kpc from the
locus of the progenitors at $\simeq 50$ Myr after the SN.
At that time, the gas density does not reach $10^3 \ \percc$ in the halo hosting the progenitors.
Before the self-enrichment, the central density of some neighbouring haloes (dubbed `1A' and `2E') 
exceeds $\sim 10^5 \ \percc$ by the halo mergers
although we do not consider molecular cooling at that regions.
We terminate the simulation at $\simeq 50 $ Myr after the SN explosions 
because the computational timestep becomes shorter and 
because star formation activity would begin in the neighbouring haloes.

\begin{figure*}
\includegraphics[width=18cm]{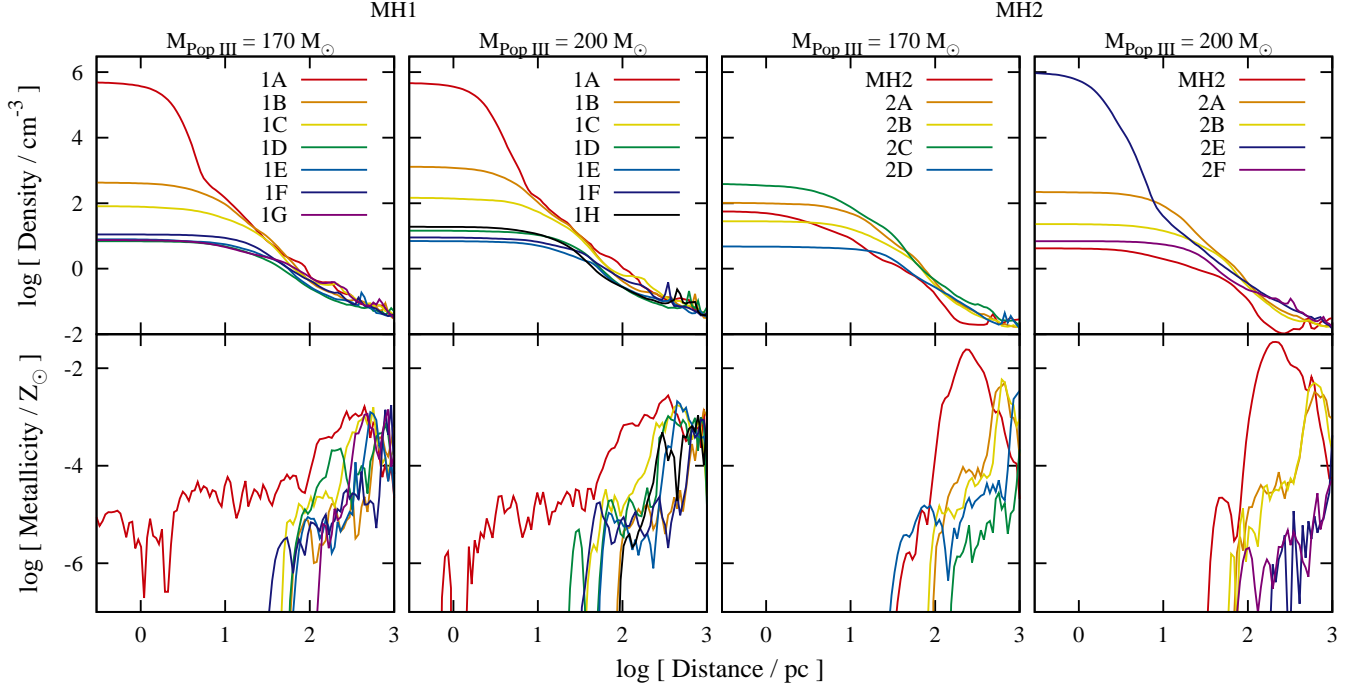}
\caption{
Density (top) and metallicity (bottom) distributions
as a function of distance from the center of the halo having hosted the Pop III star
and the enriched neighbouring haloes for PISNe in low-mass haloes (MH1 and MH2).
The labels of haloes are the same as the ones indicated in Figure \ref{fig:snapshot_ext} for each halo and $\MPopIII$.
}
\label{fig:rnz_ext}
\end{figure*}

Within 1.5 kpc from the central halo, several neighbouring haloes reside.
To study whether the neighbouring haloes are enriched or not, we pick up the ones
with non-zero metallicities within the virial radii.
The properties of haloes are listed in Table \ref{tab:results_ext}, and their loci
are shown in Figure \ref{fig:snapshot_ext}.
The configurations of these haloes are different between for {\tt P170} and {\tt P200} because
$\tlife$ and elapse time after SNe are different.
The bottom panels of Figure \ref{fig:rnz_ext} show the metallicity profile.
In most cases, the gas with metals is dammed at the regions with density $\gtrsim 10 \ \percc$
and does not reach the central densest region of the neighbouring haloes
as \citet{Chen16} reported.
Only {\tt MH1-P170}, metals are penetrated into the center of a neighbouring halo 1A,
which is formed through mergers of some haloes.
By the merger of one of the progenitors which has been superficially polluted,
the metals are incorporated into the central region.
Still, the average metallicity within the Jeans radius of 2.4 pc is $6.4\E{-6} \ \Zsun$, which
is below the critical metallicity ${\rm [Fe/H]}_{\rm cr}$, and low-mass stars are unlikely to be formed.

\begin{figure}
\includegraphics[width=8cm]{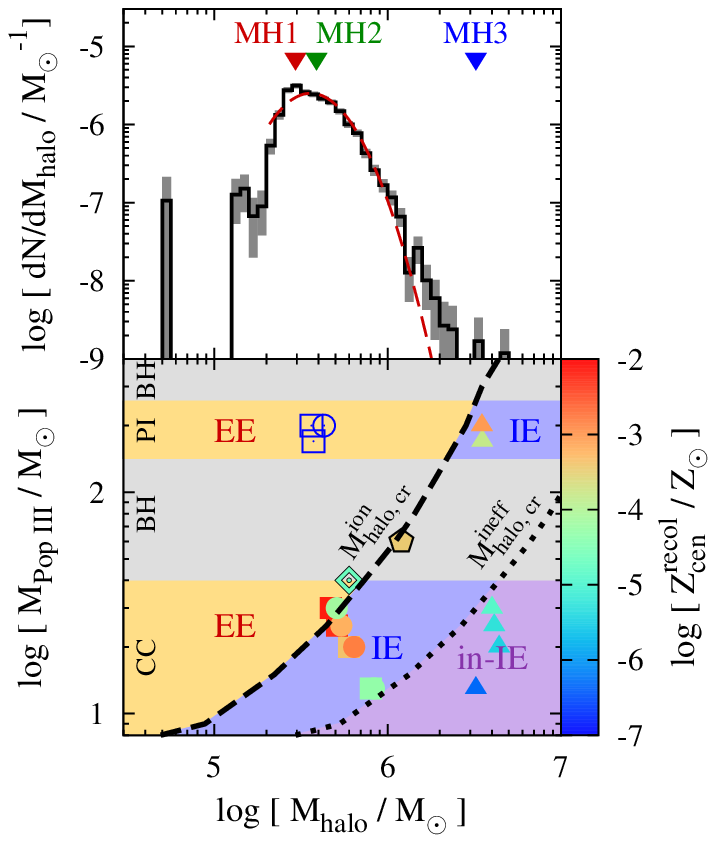}
\caption{
{\bf Top}: Mass distribution of minihaloes obtained by cosmological simulations 
by \citet{Hirano15} with Poisson errors (black solid curve).
It is fitted by a log-normal function with the mean $\mu = 12.96$ and standard deviation $\sigma = 0.41$
(red dashed curve).
The triangles indicate $\Mhalo ^{\rm col}$ of MH1, MH2, and MH3.
{\bf Bottom}: Regions on the $\MPopIII$-$\Mhalo$ plane coloured by yellow, blue, and purple indicate the parameter spaces
where the external enrichment (EE), internal enrichment (IE), and 
ineffective IE (in-IE; see Section \ref{sec:numfrac}) are realized, respectively. 
The enrichment occurs in the range of $\MPopIII$ exploding 
as CCSNe (indicated by `CC') and PISNe (`PI') while
not in the grey-coloured region with a label `BH' \citep{Heger02}.
The squares, circles, and triangles indicate $\MPopIII$ and $\Mhalo ^{\rm recol}$ 
in MH1, MH2, and MH3, respectively.
We also plot the results of \citet[][diamond]{Smith15} and
\citet[][pentagon]{Ritter16}.
The filled symbols indicate the models with IE, and 
are color-coded according to the metallicity in the recollapsing regions. 
The open symbols indicate the models with EE.
The thin and thick dotted curves represent the critical halo masses 
$M_{\rm halo, cr}^{\rm ion}$ (equation \ref{eq:MHion}) and
$M_{\rm halo, cr}^{\rm ineff}$ (equation \ref{eq:MHineff}).
}
\label{fig:Mhalo-MpopIII}
\end{figure}

\section{Critical halo masses}
\label{sec:boundary}

\subsection{Boundary between IE and EE}
\label{sec:boundary_IE_EE}

In this section, we quantitatively define the boundary between IE and EE which is schematically 
shown by dashed line in Figure \ref{fig:sche}. 
First of all, the critical halo mass below which EE realises 
is defined so that the explosion energy $\Esn$, including the effect of radiative loss, is sufficiently above
the gas binding energy $\Ebin = G M_{\rm halo} M_{\rm pri} / \Rhalo $.
We then introduce the ratio $C = \Esn / \Ebin$.
When the H {\sc ii} region fails to expand,
$C > C_{\rm cr} = 300$ is required for the MH to be brown away by the explosion.
as confirmed by one-dimensional calculations of \citet{Kitayama05}.
On the other hand, when H {\sc ii} region expands, since the SN shocks go through the rarefied
gas, $C_{\rm cr} = 1$ is applied.
We then estimate the critical halo mass below which the ratio 
exceeds $C_{\rm cr}$ as,
\begin{equation}
M_{\rm halo, cr} ^{\rm blow}
=
4.07\E{6} \ \Msun
C_{\rm cr} ^{-3/5}
\left( \frac{\Esn}{1.0\E{51} \ {\rm erg}} \right) ^{3/5},
\label{eq:Mcrit_blow}
\end{equation}
at $z_{\rm recol} \simeq 20$ with matter to baryon mass ratio 6.

Whether the gas has been diluted by the preceding photoionization
is discribed as the critical halo mass 
above which the Str{\"o}mgrem radius fails to exceed the radius $r_{\rm D} = v_{\rm D} \tlife$ 
of a D-type ionization front (I-front) within $\tlife$ to expand as an R-type front:
\begin{eqnarray}
M_{\rm halo, cr}^{\rm ion} &=&
2.5\E{6} \ \Msun
\left( \frac{v_{\rm D}}{20 \ {\rm km \ s^{-1}}} \right) ^{3/4} \nonumber \\ && \times
\left( \frac{\tlife}{5 \ {\rm Myr}} \right) ^{3/4}
\left( \frac{\QH}{10^{49} \ {\rm s^{-1}}} \right) ^{3/4}
\left( \frac{1+z_{\rm col}}{20} \right) ^{-3/2}\label{eq:MHion}
\end{eqnarray}
for the initial radial profile of clouds $\propto r^{-2}$, where $v_{\rm D}$ is 
the velocity of the D-type I-front \citep{Kitayama04}.
In Figure \ref{fig:Mhalo-MpopIII},
the dashed curve shows $M_{\rm halo, cr}^{\rm ion}$ with $\tlife$ and $\QH$ 
of the results of \citet{Schaerer02}.

Then, the condition for EE is written as
\begin{eqnarray}
M_{\rm halo}
<  \begin{cases}
M_{\rm halo, cr}^{\rm blow} (C_{\rm cr} = 1) 
\ \ \ {\rm for \ M_{\rm halo} < M_{\rm halo, cr}^{\rm ion}} \\
M_{\rm halo, cr}^{\rm blow} (C_{\rm cr} = 300) 
\ \ \ {\rm for \ M_{\rm halo} > M_{\rm halo, cr}^{\rm ion}}
  \end{cases}
\end{eqnarray}
When H {\sc ii} region can expand ($M_{\rm halo} < M_{\rm halo, cr}^{\rm ion}$;
top-left of Figure \ref{fig:Mhalo-MpopIII}),
MHs with $M_{\rm halo} < M_{\rm halo, cr}^{\rm blow} (C_{\rm cr} = 1) =
 4.07\E{6} \ \Msun$ (CCSN) and 
$2.74\E{7} \ \Msun$ (PISN),
that is, the all regions with $M_{\rm halo} < M_{\rm halo, cr}^{\rm ion}$ undergo EE.
When H {\sc ii} region fails to expand  ($M_{\rm halo} > M_{\rm halo, cr}^{\rm ion}$), MHs with $M_{\rm halo} <
 1.33\E{5} \ \Msun$ (CCSN) and 
$8.94\E{5} \ \Msun$ (PISN)
undergo EE.
Although a tiny fraction of the region with $M_{\rm halo} < M_{\rm halo, cr}^{\rm ion}$
is favoured for EE for CCSN, let us define the condition for EE is $M_{\rm halo} < M_{\rm halo, cr}^{\rm ion}$
for simplification.
For convenience, we present the fitting formula of the boundary as
\begin{eqnarray}
M_{\rm halo, cr}^{\rm ion} &\simeq&
6\E{5} \ \Msun \nonumber \\ && \times
\left\{
\left( \frac{\MPopIII}{25 \ \Msun} \right)
\left[
1-\exp
\left( -\frac{\MPopIII}{25 \ \Msun} \right) 
\right]
\right\} ^{3/4}
\label{eq:MHion_fit}
\end{eqnarray}
valid for $\MPopIII = 10$--$200 \ \Msun$ and $z_{\rm col} = 20$--$30$.

Note that {\tt MH1-C30} and {\tt MH2-C30} are marginally judged to be the models with EE
by this criterion.
The estimation of the critical halo mass is still simplified.
More simulations with various initial parameters can make the estimation more accurate.

In Figure \ref{fig:Mhalo-MpopIII}, we also plot the results of previous works.
Their results agree with our criterion.
\citet{Smith15} find that EE occurs with the initial halo mass $M_{\rm halo} = 5\E{5} \ \Msun$
and $\MPopIII = 40 \ \Msun$, and $Z_{\rm cen} = 2\E{-5} \ \Zsun$.
\citet{Ritter16} show that IE occurs with $M_{\rm halo} = 1\E{6} \ \Msun$
and $\MPopIII = 60 \ \Msun$, and resulting metallicity is $2$--$5\E{-4} \ \Zsun$.
To consider the accretion of pristine gas into MHs, we plot them on $\Mhalo$ 1.2 times 
larger than their initial values.
These parameters are slightly above and below $M_{\rm halo, cr}^{\rm ion}$, respectively.

\subsection{Condition for ineffective IE}
\label{sec:boundary_IE-a_IE-b}

We also define the condition for the ineffective IE presented in Section \ref{sec:summary_dependency}.
In Figure \ref{fig:sche}, the region of ineffective IE exists lower right corner, 
where a less massive Pop III star explodes in a massive minihalo.
As the necessary condition for this mode, SN remnants should undergo 
off-central recollapse, where H {\sc ii} region can hardly break the Pop III star-forming 
clouds, and even continue to collapse with ineffectively accreting metals.
The critical condition (dashed curve in Figure \ref{fig:Mhalo-MpopIII}) is consistent with 
the boundary between the central recollapse and off-central recollapse seen in our simulations. 
The parameters ($\Mhalo$ and $\MPopIII$) for {\tt MH1-C13} and {\tt C20}, where 
the off-central recollapse occurs, are below the dotted line.
The legend's colour shows the metalliciites in the recollapsing regions in the simulations.
The boundary separates the parameters which gives the metallicities above and below 
${\rm [Fe/H]}_{\rm cen}^{\rm recol} \sim -3$.
We define the ineffective IE as the cases with 
${\rm [Fe/H]}_{\rm cen}^{\rm recol} \leq -5$.
For smaller metallicity boundary, larger $\Mhalo$ is permitted because it can 
retain more dense pristine clouds during MS of Pop III stars into which metals
more hardly penetrate.
Multiplying $M_{\rm halo, cr}^{\rm ion}$ with a factor of six can reproduce
the critical halo mass $M_{\rm halo, cr}^{\rm ineff}$ above which ineffective IE occurs
as indicated by the dotted curve:
\begin{eqnarray}
M_{\rm halo, cr}^{\rm ineff} &=&
4\E{6} \ \Msun \nonumber \\ && \times
\left\{
\left( \frac{\MPopIII}{25 \ \Msun} \right)
\left[
1-\exp
\left( -\frac{\MPopIII}{25 \ \Msun} \right) 
\right]
\right\} ^{3/4}.
\label{eq:MHineff}
\end{eqnarray}

\subsection{Number fraction of MHs for each enrichment process}
\label{sec:numfrac}
As a next step, we estimate the number fractions of MHs where each enrichment mode occurs. 
For this purpose, we use the mass distribution of 1538 MHs hosting Pop III stars taken from a cosmological simulations \citep{Hirano15}.
The top panel of Figure \ref{fig:Mhalo-MpopIII} shows it with error bars estimated as the Poisson error.
The lower limit is determined by the minimum virial temperature $T_{\rm vir} = 1000$ K above which
primordial gas can collapse by the radiative cooling owing to hydrogen molecules while
the upper limit is determined by the redshift-dependent intensity of dissociation photons from the neighbouring haloes.
In this range, the mass distribution of MHs is fitted by a log-normal function 
truncated at those masses. 
With the least squares, its mean and standard deviation are estimated as $\mu = 12.96\pm 0.02$ and $\sigma = 0.41\pm 0.02$, respectively (red dashed curve in Figure \ref{fig:Mhalo-MpopIII}).

%

Since the critical halo masses
($M_{\rm halo, cr}^{\rm ion}$ and $M_{\rm halo, cr}^{\rm ineff}$)
depend on $\MPopIII$, we should assume some 
IMF of Pop III stars:
the flat and Salpeter IMFs with slope indices $0$ and $-2.35$, respectively.
Using this halo mass function and Pop III IMF, 
the number fractions of MHs with each enrichment mode are estimated 
for PISNe and CCSNe (Table \ref{tab:PopII_fraction}).

For PISNe, our simulations suggest that almost all MHs undergo EE
in the relevant mass range by the large efficiency of photoionization and 
large explosion energy regardless of the Pop III IMF.
This shows that negligible fractions ($\sim 10^{-6}$--$10^{-5}$) of MHs create low-mass EMP stars with 
the abundance ratio of PISNe even if such massive progenitors exist.
It is consistent with the observations showing that such EMP stars have not been observed so far
as further discussed in the next section.

For CCSNe, EE is predicted to occur 42\% and 13\% of MHs for the flat
and Salpeter IMFs, respectively.
More than half of MHs are sufficiently self-enriched to form metal-poor star-forming
clouds with metallicities ${\rm [Fe/H]}_{\rm cen}^{\rm recol} > -5$.
For the flat and Salpeter IMF, only 1\% and 5\% of MHs more massive than 
$M_{\rm halo, cr}^{\rm ineff}$, thereby they undergo ineffective IE.

\begin{table}
\caption{Fractions of minihaloes for each enrichment process}
\label{tab:PopII_fraction}
\begin{tabular}{ccccc}
\hline
 SN & Enr. mode & ${\rm [Fe/H]}_{\rm cen}^{\rm recol}$ & \multicolumn{2}{c}{Fraction} \\
    &           &                                       & flat IMF    & Salpeter IMF  \\
\hline \hline                                                        
 PI & EE        & $<{\rm [Fe/H]}_{\rm cr}$              & $\simeq 1$   & $\simeq 1$   \\
    & IE        & $>{\rm [Fe/H]}_{\rm cr}$              & $4.60\E{-6}$ & $7.23\E{-6}$ \\
\hline                                                                   
 CC & EE        & $<{\rm [Fe/H]}_{\rm cr}$              & $0.423$      & $0.127$      \\
    & IE        & $>{\rm [Fe/H]}_{\rm cr}$              & $0.566$      & $0.824$      \\
    & in-IE     & $<{\rm [Fe/H]}_{\rm cr}$              & $0.011$      & $0.049$      \\
\hline
\end{tabular}
\medskip \\
Note --- (2) Enrichment mode: external enrichment (EE), internal enrichment (IE), and ineffective IE (in-IE).
See Section \ref{sec:numfrac} for the detailed descriptions.
\end{table}

\section{Comparison with the observations of EMP stars}
\label{sec:observation}

\subsection{Formation of massive Pop III exploding as PISNe}

Although we see that IE occurs for our {\tt HH-PISN} models,
such a massive halo with $\geq 3\E{6} \ \Msun$ is very rare (Figure \ref{fig:Mhalo-MpopIII}).
This indicates that Pop II stars which inherit elemental abundance ratios of
PISNe are hardly formed in the metallicity range of ${\rm [Fe/H]} > -5$ because
almost all MHs are blown away by the copious emission of ionizing photons from
massive Pop III stars and energetic explosion of PISNe even if they exist.
Although 1,401 stars with ${\rm [Fe/H]} < -2.5$ have already been observed until September 2017,
EMP stars with the elemental abundance ratio of PISNe have 
not been observed yet, or only one candidate \citep{Aoki14}.
Our results suggest that $10^5$--$10^6$ stars should be observed to find a star with a sign of PISNe.

Some researches suggest that even the existence of the progenitors is unlikely.
Some numerical works show the fragmentation of primordial clouds \citep{Bromm99} 
and accretion disk around a Pop III protostar \citep{Greif11, Stacy12, Susa14, Hosokawa16, Hirano17}.
Also, it has been suggested that even with the Salpeter IMF
the inefficient sampling of stellar mass fails to create such massive stars exploding as PISNe, 
because MHs are expected to have smaller star formation efficiencies 
due to their lack of coolant \citep{deBennassuti14, deBennassuti17}.

\begin{table*}
\begin{minipage}{15cm}
\caption{CEMP stars}
\label{tab:CEMP_fraction}
\begin{tabular}{cccccccccc}
\hline
 &
\multicolumn{2}{l}{Observation} & \multicolumn{5}{l}{Model} & \multicolumn{2}{l}{This work} \\
Star    & $\abFe{C}_{\rm obs}$ & $\abH{Fe}_{\rm obs}$ & Ref.                    & $\MPopIII$ & $\Esn$             & $\Mcut$   & $M_{\rm Fe} ^{\rm ej}$ & Halo & $\abH{Fe}_{\rm cen}^{\rm recol}$ \\
        &                      &                      &                         & [$\Msun$]  & [$10^{51} \ \erg$] & [$\Msun$] & [$\Msun$]              &      &                                  \\
\hline \hline
 HE0557 & $1.76$               & $-4.79$              & \citetalias{Ishigaki14} & 25         & 1                  &   1.7     & $2.2\E{-3}$            & {\bf MH1}  & ${\bf -4.34}$              \\
        &                      &                      &                         &            &                    &           &                        & MH2  & $-5.29$                          \\
        &                      &                      &                         &            &                    &           &                        & MH3  & $-7.34$                          \\
        &                      &                      & \citetalias{Tominaga14} & 25         & 5.0                &   1.78    & $6.80\E{-3}$           & MH1  & $-3.85$                          \\
        &                      &                      &                         &            &                    &           &                        & {\bf MH2}  & ${\bf -4.80}$              \\
        &                      &                      &                         &            &                    &           &                        & MH3  & $-6.85$                          \\
        &                      &                      & \citetalias{Marassi15}  & 20         & 1.0                &   5       & $2.82\E{-5}$           & MH1  & $-7.16$                          \\
        &                      &                      &                         &            &                    &           &                        & MH2  & $-6.61$                          \\
        &                      &                      &                         &            &                    &           &                        & MH3  & $-9.24$                          \\
\hline
 HE0107 & $4.12$               & $-5.44$              & \citetalias{Ishigaki14} & 25         & 1                  &   1.7     & $1.4\E{-4} $           & {\bf MH1}  & ${\bf -5.53}$              \\
        &                      &                      &                         &            &                    &           &                        & MH2  & $-6.48$                          \\
        &                      &                      &                         &            &                    &           &                        & MH3  & $-8.54$                          \\
        &                      &                      & \citetalias{Tominaga14} & 25         & 5.0                &   1.90    & $8.02\E{-5}$           & {\bf MH1}  & ${\bf -5.78}$              \\
        &                      &                      &                         &            &                    &           &                        & MH2  & $-6.73$                          \\
        &                      &                      &                         &            &                    &           &                        & MH3  & $-8.78$                          \\
\hline
 HE1327 & $4.18$               & $-5.71$              & \citetalias{Ishigaki14} & 25         & 1                  &   1.7     & $1.1\E{-5} $           & MH1  & $-6.64$                          \\
        &                      &                      &                         &            &                    &           &                        & MH2  & $-7.59$                          \\
        &                      &                      &                         &            &                    &           &                        & MH3  & $-9.64$                          \\
        &                      &                      & \citetalias{Tominaga14} & 25         & 0.72               &   1.69    & $1.53\E{-5}$           & MH1  & $-6.50$                          \\
        &                      &                      &                         &            &                    &           &                        & MH2  & $-7.45$                          \\
        &                      &                      &                         &            &                    &           &                        & MH3  & $-9.50$                          \\
        &                      &                      & \citetalias{Marassi15}  & 30         & 1.6                &   5       & $1.58\E{-5}$           & MH1  & $-6.66$                          \\
        &                      &                      &                         &            &                    &           &                        & MH2  & $-8.63$                          \\
        &                      &                      &                         &            &                    &           &                        & MH3  & $-9.37$                          \\
\hline
        & $\abFe{C}_{\rm obs}$ & $\abH{C}_{\rm obs}$  &                         &            &                    &           & $M_{\rm C} ^{\rm ej}$  &      & $\abH{C}_{\rm cen}^{\rm recol}$  \\
        &                      &                      &                         &            &                    &           & [$\Msun$]              &      &                                  \\
\hline
 SM0313 & $>4.9$               &  $-2.6$              & \citetalias{Ishigaki14} & 25         & 1                  &   2.0     & $2.88\E{-1} $          & {\bf MH1}  & ${\bf -2.49}$              \\
        &                      &                      &                         &            &                    &           &                        & MH2  & $-3.43$                          \\
        &                      &                      &                         &            &                    &           &                        & MH3  & $-5.48$                          \\
\hline
\end{tabular}
\medskip \\
Note --- 
The observed data $\abFe{C}_{\rm obs}$ and $\abH{Fe}_{\rm obs}$ for the CE-EMP stars
are taken from SAGA database and references therein.
Using the Fe mass $M_{\rm Fe}^{\rm ej}$ ejected from the progenitors modeled by
\citet[][\citetalias{Ishigaki14}]{Ishigaki14},
\citet[][\citetalias{Tominaga14}]{Tominaga14}, and
\citet[][\citetalias{Marassi15}]{Marassi15},
we estimate the iron abundance $\abH{Fe}_{\rm en}^{\rm recol}$ in the recollapsing region
when each progenitor explodes in MH1, MH2, and MH3.
The MH models in which $\abH{Fe}_{\rm en}^{\rm recol}$ is close to $\abH{Fe}_{\rm obs}$ 
within 0.5 dex are written in bold.
\end{minipage}
\end{table*}

\subsection{C-enhanced EMP stars}

We have calculated the metallicity in the recollapsing region for CCSNe which produce
the C-normal abundance ratio.
For stars with $\abH{Fe} < -3$, the fraction of CE-EMP stars increase as [Fe/H] decreases \citep{Beers05}.
In order to  reproduce the elemental abundance of CE-EMP stars, the model of faint 
SNe (FSNe) is often employed.
In the mixing-fallback scenario, 
the inner layer of ejecta enriched with Fe-peak elements sink into central compact objects 
while the outer C-rich layer is ejected.
Although the models of a single FSN can well reproduce abundance patterns of CE-EMP stars,
it has not been numerically shown that the metallicity of the gas enriched by a FSN can reproduce 
the absolute metallicity, or Fe abundance $\abH{Fe}$.

Our results can be applied to 
general SN models with the common progenitor mass $\MPopIII$ and explosion energy $\Esn$, 
which determine the density distribution of the ambient gas.
Since we do not calculate the back reaction of the metal distribution to the hydrodynamic motions of gas 
such as the additional radiative cooling owing to metal and dust,
we can reproduce the spatial distribution of metal ejected from any SN models by just scaling the total metal mass.
We here employ models tailored to reproduce the elemental abundances of CE-EMP stars,
$\rm HE0557$-$4840$ (HE0557),
$\rm HE0107$-$5240$ (HE0107),
$\rm HE1327$-$2326$ (HE1327), and
$\rm SMSS J0313$-$6708$ (SM0313),
presented by 
\citet[][\citetalias{Ishigaki14}]{Ishigaki14},
\citet[][\citetalias{Tominaga14}]{Tominaga14}, and
\citet[][\citetalias{Marassi15}]{Marassi15}.
To compare with observations, we estimate ${\rm [Fe/H]}_{\rm cen}^{\rm recol}$ 
from the iron mass $M_{\rm Fe}^{\rm ej}$ of each SN model with equation (\ref{fig:Z2Fe}).
Since the efficiency of metal dispersion varies among the MHs, we estimate
${\rm [Fe/H]}_{\rm cen}^{\rm recol}$ for each MH as shown in Table \ref{tab:CEMP_fraction}.
The MH models which can reproduce ${\rm [Fe/H]}_{\rm obs}$ within 0.5 dex are written in bold.

For the star HE0557 with a moderate C-enhancement $\abFe{C} = 1.76$,
the resulting metallicities for the progenitor model of \citetalias{Ishigaki14} 
for MH1 is consistent with the observed metallicity $\abH{Fe} _{\rm obs} = -4.79$.
The model of \citetalias{Tominaga14} for MH2 can explain the formation of HE0557.
Note that $\Esn$ in their model is $5.0\E{51}$ erg while $\Esn = 1.0\E{51}$ erg 
in our simulations.
Simulations with $\Esn = 5.0\E{51}$ erg would result in smaller metallicities because
the metals would be brown away to larger distances.
For \citetalias{Marassi15}, the metallicity $\abH{Fe}_{\rm cen}^{\rm recol}$ is smaller than
the observed abundance because of the larger mass-cut $\Mcut = 5 \ \Msun$ and accordingly
smaller ejected metal mass.
For the massive halo MH3, where ineffective IE takes place, the metallicities are smaller 
than the observed one by more than three orders of magnitude for every progenitor model.
This indicates that IE from FSN hosted by MH1 and MH2 with normal halo mass 
$\Mhalo \sim 3\E{5} \ \Msun$ can explain the formation of the C-moderate star.

For the star HE0107, $M_{\rm Fe}^{\rm ej}$ and $\abH{Fe}_{\rm cen}^{\rm recol}$
are consistent among the literatures.
Although \citetalias{Marassi15} present the progenitor model,
their best-fit model is for $\MPopIII = 35 \ \Msun$ with which we do not simulate the process of metal dispersion.
In the models of \citetalias{Ishigaki14} and \citetalias{Tominaga14} for MH1, 
$\abH{Fe}_{\rm en}^{\rm recol} = -5.53$ and $-5.78$, respectively, are slightly
smaller than $\abH{Fe}_{\rm obs}$ by a factor of 0.1--0.3 dex.\footnote{Take care 
that the iron abundance ${\rm [Fe/H]}_{\rm cen}^{\rm recol} < {\rm [Fe/H]}_{\rm cr}$ 
does not violate the critical condition for the formation of low-mass stars in the 
CE-EMP regime.
Instead of silicate grains, carbon grains are dominant species and gives the 
critical C abundance as $-4.65 < {\rm [C/H]}_{\rm cr} < -3.01$ \citep{Marassi14}.}
The predicted metallicity for MH2 is smaller than for MH1 
because a pristine cloud falls into the central halo before the shocked shell accretes again.
As for HE0557, the metallicity for MH3 is further smaller than MH1 by three orders of magnitude.
Let us thus conclude that IE from a FSN hosted by a less-massive halo without sub-structures 
is most likely to explain the formation of the star HE0107.

For HE1327, the tendency of metallicity among MHs is the same as for HE0107.
However, the predicted metallicity is smaller than the observed one by a factor of at least
0.8--0.9 dex for MH1.
The FSN scenario generally predicts the smaller mass of ejected metal relative to 
normal CCSNe due to the fallback of material into the core.
To create the star-forming region simultaneously satisfying the large C-enhancement $\abFe{C} = 4.18$
and metallicity $\abH{Fe} = -5.71$, multiple SN explosions including FSN would be required.

For SM0313, \citet{Keller14} measure only the upper limit of its iron abundance.
Therefore, we compare the carbon abundance predicted from our simulations with the one constrained by the observation.
\citetalias{Ishigaki14} and \citetalias{Marassi15} present the progenitor models which can reproduce
its elemental abundance ratio.
The best-fit model by \citetalias{Marassi15} is with $\Esn = 80 \ \Msun$, for which we here do not calculate.
The model with $\MPopIII = 25 \ \Msun$ of \citetalias{Ishigaki14} predicts the carbon 
mass of $M_{\rm C}^{\rm ej} = 0.288 \ \Msun$.
For MH1, we can predict the carbon abundance in the recollapsing clouds to be 
$\abH{C}_{\rm cen}^{\rm recol} = -2.49$,
which is consistent with the observation $\abH{C}_{\rm obs} = -2.6$ with a 3D-LTE model.
For MH2, the C abundance is smaller as $\abH{C}_{\rm cen}^{\rm recol} = -3.43$ 
because the pristine clouds merge with the central cloud.
It is likely that IE from a single FSN hosted by isolated less-massive halo
can explain the formation of the most iron-poor star observed. 

\begin{figure*}
\includegraphics[width=12cm]{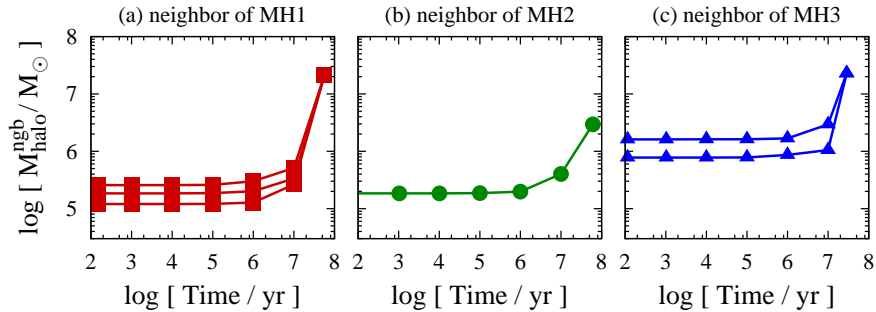}
\caption{
Mass of neighbouring haloes $\Mhalo ^{\rm ngb}$ 
around (a) MH1, (b) MH2, and (c) MH3 
as a function of time after SN explosion.
}
\label{fig:tMhalo_neighbor}
\end{figure*}

\subsection{Critical metallicity}
\label{sec:critical_metallicity}

For {\tt HH-CCSN}, the metallicity resulting from the off-central 
recollapse is very small, less than ${\rm [Fe/H]} < -5$ (the triangle symbols in
Figure \ref{fig:C-Fe}).
For FSNe exploding in MH3, the carbon fraction becomes ${\rm [C/H] < -3}$
(Table \ref{tab:CEMP_fraction}).
Figure \ref{fig:C-Fe} shows that EMP stars with this metallicity 
range ($10^{\rm [C/H]-2.30} + 10^{\rm [Fe/H]} < 10^{-5.07}$) have 
so far not been found (gray shaded region).
From the theoretical point of view, clouds in such a metallicity range collapse to form a single massive objects 
due to the lack of cooling efficiency of dust grains \citep{Omukai00, Schneider03, Chiaki17}.
On the other hand, some researchers report that the fragmentation of the
accretion disk around a Pop III protostellar core leads to low-mass star
formation below the critical metallicity.
Consequently, the clouds with ${\rm [Fe/H]} < -5$ might have similar properties of 
fragmentation process, since its thermal evolution is almost identical to the primordial one.

By EE and ineffective IE modes, metallicities in the recollapsing regions do not reach
the critical metallciity.
Our simulations predict the
number fraction of MHs which undergo EE and ineffective IE as 18--43\% for CCSNe 
(Section \ref{sec:numfrac}).
Suppose that the slightly enriched clouds in these MHs collapse to low-mass objects through 
the fragmentation,
then the number fraction of EMP star would also become 18--43\% for the simple estimation.
From the number of EMP stars ever observed (1,401), the number of stars below the critical 
metallicity would be 252--602, which can not explain the current observations.
Therefore, the scenario of dust-driven fragmentation would be favored to the current observation.
We should note that the estimation may be oversimplified because we assume that 
the IMF of Pop III is uniform in all mass range of MHs and that the number of 
fragments is uniform in all metallicity range.

\section{Discussion}
\label{sec:discussion}

\subsection{Enrichment efficiency of MHs}
Although we have so far estimated the averaged metallicities
of the recollapsing gas clouds within their Jeans length and 
of EMP stars which collapse earliest, 
the succeeding EMP star formation will take place in the whole region of MHs.
We in this section estimate the average metallicity and the efficiency of metal enrichment of each MH
as indicated by the dashed lines in Figure \ref{fig:MpopIII-Zcen_int} and the last column of Table \ref{tab:results}.

We introduce the mass ratio of metal and primordial gas particles 
within the virial radius at $\trecol$ to the initial as
\begin{eqnarray}
f_{\rm halo} ^{\rm met} &=& \frac{M_{\rm met,halo} ^{\rm recol}}{M_{\rm met}^{\rm ej}}, \\
f_{\rm halo} ^{\rm pri} &=& \frac{M_{\rm pri,halo} ^{\rm recol}}{M_{\rm pri}^{\rm col}}.
\end{eqnarray}
We only consider metal and gas particles trapped by the DM potential well with the 
radial velocity $v_{\rm rad}$ smaller than the escape velocity 
$v_{\rm esc} = (2 G \Mhalo / \Rhalo )^{1/2}$.
These factors are shown in the seventh to eighth columns of Table \ref{tab:results}. 

The simple estimation of the metallicity in a MH is the mass ratio
$M_{\rm met, halo}^{\rm recol} / M_{\rm pri,halo} ^{\rm recol}$.
The factor $f_{\rm halo} ^{\rm met} / f_{\rm halo} ^{\rm pri}$ gives the correction
for it.
For example, $M_{\rm met, halo}^{\rm recol} / M_{\rm pri,halo} ^{\rm recol} = 9.18\E{-3} \ \Zsun$
for {\tt MH1-C30}, but the 60.2\% of metal is dispersed into the voids and the MH accretes the primordial
gas with 2.62 times its initial mass. 
The dilution of metal by the factor of $f_{\rm halo} ^{\rm met} / f_{\rm halo} ^{\rm pri}= 0.152$ results in the eventual metallicity
$1.39\E{-3} \ \Zsun$. \\

\noindent{\it CCSNe in low-mass MHs} \\

For {\tt LH-CCSN}, the metallicity range is $\sim 10^{-5}$--$10^{-3} \ \Zsun$, 
comparable to $Z_{\rm cen}^{\rm recol}$ in the center of the recollapsing regions.
The dashed lines in Figure \ref{fig:MpopIII-Zcen_int} shows that
the dependency of the metallicity on $\MPopIII$ is mitigated relative to the solid lines
by averaging the metallicity in the wider region 
beyond the offset between the metal and primordial gas distributions for ineffective IE.
Since the accretion of primordial gas overcomes its evacuation by SN explosion, 
the primordial gas mass increases by a factor of 44.2--117.2.
Whereas, about a half (0.254--0.701) of metals remain in the virial radius.
Consequently, the resulting metallicity is smaller than the simple estimation by a factor of $f_{\rm halo}^{\rm met}/f_{\rm halo}^{\rm pri} \sim 0.1$.
For {\tt MH2-CCSN}, $f_{\rm halo}^{\rm pri}$ is larger than for {\tt MH1-CCSN} because of the halo merger. \\

\noindent{\it CCSNe in massive MHs} \\

For {\tt HH-CCSN}, since a large fraction of metal does not enter the central clouds 
but remains within the virial radius (Figure \ref{fig:rnz_int}), the metallicity 
range is $3.63\E{-5}$--$2.43\E{-4} \ \Zsun$, slightly above the critical metallicity
(Figure \ref{fig:MpopIII-Zcen_int}).
Both $f_{\rm halo}^{\rm met}$ and $f_{\rm halo}^{\rm pri}$ are larger than those 
for {\tt LH-CCSN} by the larger accretion rate and by the smaller $\trecol$. \\

\noindent{\it PISNe in massive MHs} \\

For {\tt HH-PISN}, both $f_{\rm halo}^{\rm met}$ and $f_{\rm halo}^{\rm pri}$ is 
comparable those for {\tt HH-CCSN}.
However, the mass $M_{\rm met}^{\rm ej}$ is larger by a factor of ten (Table \ref{tab:PopIIIstars}), 
the metallicity is also larger by the same factor.
For {\tt MH3-P200}, the metallicity is smaller than {\tt P170} because a neighbouring halo is merged, and 
the dense primordial envelope spreads around the enriched cloud. \\

\subsection{Caveats}
Our method can uncover the various nature of metal enrichment occurring in the transition from
first-generation metal-free Pop III stars to Pop II star formation.
Some caveats still remain in our methodology.

\subsubsection{Star formation in neighbouring haloes}
In our simulations, gas contraction is prohibited by switching off molecular cooling
in the other haloes than the central one only whose stellar feedbacks interests us in this work.
In reality, feedbacks from stars formed in neighbouring haloes can affect the propagation of shocks from the central star.
Also, in some recollapsing regions, the metallicity will be the sum of the contributions from more than one halo,
and the elemental abundance ratio will be the superposition of them.
In this section we discuss the efficiency of the feedback from stars hosted by the neighbouring haloes,
following the evolution of their mass $\Mhalo ^{\rm ngb}$ in the case of  {\tt P200}, where
we can follow the longest growth history of haloes.

The halo 1A (see Figure \ref{fig:snapshot_ext}) becomes the closest to MH1.
It is formed through the merger of three haloes and accretion of the intergalactic gas.
Figure \ref{fig:tMhalo_neighbor}(a) shows the temporal evolution of the masses
of the progenitors and 1A.
The progenitors contain $1.20$, $1.84$, and $2.57\E{5} \ \Msun$ within $\sim 1$ Myr after the
SN explosion in MH1, and they increase by a factor of two at 10 Myr.
They reside at the distance $D \simeq 400$ pc from MH1.
At $\simeq 50$ Myr, eventually, the mass and distance of 1A becomes $2.14\E{7} \ \Msun$ and $\simeq 200$ pc, respectively.
For CCSNe, since the SN shocks should go beyond the dense filament to reach MH1,
they are unlikely to affect the hydrodynamic evolution of shocks from MH1.
For PISNe, although the shocks and metal will reach MH1,
metal can not penetrate the central clouds as we have seen above.
Since the mass evolution of the neighbouring halo of MH2 is similar to that of MH1
(Figure \ref{fig:tMhalo_neighbor}b), we can conclude that the external explosion
around MH2 neither likely affect the propagation of shock and metal from MH2.

For massive halo (MH3), the neighbouring haloes are also massive with $7.76\E{5}$
and $1.61\E{6} \ \Msun$ in the initial state.
Thus, the feedbacks from the stars in these haloes can not affect the hydrodynamics
of the shocks from MH3 through the filaments connecting them to MH3.
However, these haloes eventually merge with MH3 at $\simeq 10^7$ Myr after the SN.
When the stars in the neighbouring haloes explode in the vicinity of MH3,
the shock will affect and metal is mixed in the recollapsing region.
We will examine this by permitting star formation in all haloes in forthcoming papers.

\subsubsection{Multiple star formation in a halo}
We assume that each MH hosts a single Pop III star.
However, the number of Pop III stars which are hosted by a MH is still uncertain.
Multiple star formation in a MH through the fragmentation of accretion disk around the primary protostar
has been reported by some numerical works 
\citep{Bromm99, Clark11, Greif11, Stacy12, Susa13, Susa14, Hosokawa16, Hirano17}.
Majority of those fragments seem to fall in the range of CCSN.
On the other hand, the explosion energy of PISNe is larger than that of CCSNe by a factor of $\sim 30$. 
Hence if the number of fragments is $\gtrsim 30$,
the total explosion energy becomes larger than that of a single PISN.

However, it has been reported that the number of fragments less than 30
thereby the multiple CCSNe would be less energetic than the PISN case investigated 
in the present simulations. 
Thus the dynamics of the ejecta of the multiple CCSNe  would be in between the CCSN 
and PISN cases shown in the present paper.  It also will increase the amount of 
metals by some factors. 
These issues will be addressed in the forthcoming papers.

\begin{figure}
\includegraphics[width=8cm]{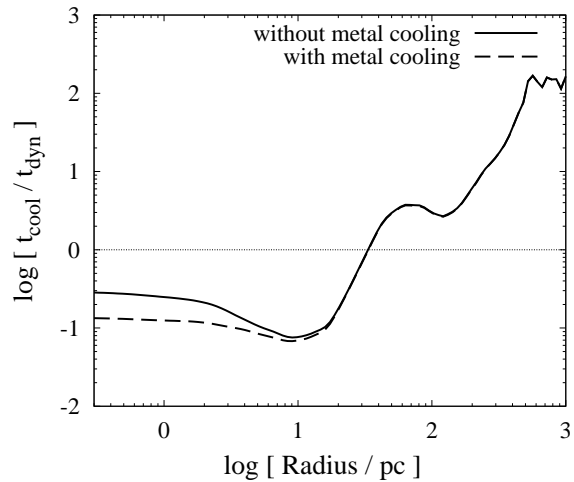}
\caption{
Ratio of cooling time $t_{\rm cool}$ to dynamical time-scale $t_{\rm dun}$ 
as a function of the distance from the most dense point in
the recollapsing region with a Pop III mass $\MPopIII = 30 \ \Msun$ hosted by MH1.
The solid curve shows the result only with cooling by primordial species, which is used in our
simulations.
The dashed curve shows the result by adding the contribution of metal species to gas cooling
as a post-process.
Here, we estimate the local free-fall time as $t_{\rm dyn}$.
}
\label{fig:rtcool_MH001_M03P200_300}
\end{figure}

\subsubsection{Resolution}
Metals mix into the pristine gas mainly due to the turbulence driven by fluid instabilities around shocks.
Since the size of eddies cascades into smaller scales, 
it is ideally required to resolve down to the scale 
at which the eddies are dissipated to see the full mixing processes.
In our simulations, the minimum spacial resolution is 1.3 pc (equation \ref{eq:resolution})
because of the large simulation region.
In order to investigate the wide range of the parameters, we can not perform convergence tests.

Instead, we compare our results with the recent work \citep{Chen16}:
high-resolution two-dimensional simulations of the region 
where a SN shock hits a neighbouring pristine cloud with the minimum resolution of 0.015 pc.
The separation of the target cloud and SN is 250 pc 
both in their work and in our {\tt LH-PISN} models (Table \ref{tab:results_ext}).
They show that SN shocks drive the Rayleigh-Taylor (RT) instability at the wall of the target cloud, 
but metals fails to incorporate into the clouds in  most cases.
The penetration of metals halt at the position 30 pc away from the cloud center (figure 9 in their paper) 
as seen also in our results (Figure \ref{fig:rnz_ext}).
Even with our spatial resolution coarser, 
the behavior of metal mixing is consistent with their results.

The resolution also affects the cooling efficiency by metals in a dense snowplough shell
and recollaising region.
We ignore metal cooling in the enriched gas.
In the simulations of cosmological star formation history by \citet{Schaye10},
the global star formation rate is underestimated in the run without metal cooling.
\citet{Ritter16} shows that fragments insufficiently enriched by a SN undergo loitering for 10 Myr 
while cooling time in a clump with $\gtrsim 10^{-3} \ \Zsun$ becomes smaller than the dynamical time-scale
due to C and O fine-structure coolings to start run-away collapse. 
Figure \ref{fig:rtcool_MH001_M03P200_300} shows the ratio of cooling time $t_{\rm cool}$ to dynamical time $t_{\rm dyn}$
as a function of distance from the collapse center in our simulation with {\tt MH1-C30},
where the metallicity in the recollapsing center becomes $6.62\E{-3} \ \Zsun$.
The ratio becomes smaller for the case with metal cooling (dashed curve) than only 
with contribution of primordial species (solid curve) by a factor of two.
In that case, cloud collapse becomes faster, and efficiency of metal mixing would become lower.
We will confirm this in the forthcoming paper.

\section{Conclusion}
\label{sec:conclusion}

By the numerical simulations for the ionization and SN feedbacks, 
we estimate the metallicities in the recollapsing region 
which will be Pop II star-forming clouds or continue to be in the Pop III star formation regime.
Thanks to the simulations with a range of initial parameters, $\Mhalo$ and $\MPopIII$, and 
the initial gas distributions with various substructures, 
we find that a few enrichment modes can occur, and the metallicity range accordingly varies.
We have found that

\begin{itemize}
\item there are two general enrichment processes by Pop III SNe: internal enrichment (IE) and external enrichment (EE).
They exclusively occurs depending on $\Mhalo$ and $\MPopIII$ ($\Esn$).

\item for CCSNe, a non-negligible fraction (13--42\%) of MHs undergo EE to form
recollapsing clouds with ${\rm [Fe/H]}_{\rm cen}^{\rm recol} < -5$.
So far no CN-EMP stars with the metallicity range are observed, indicating that the 
inefficient dust cooling prevents low-mass stars from being formed.

\item for PISNe, IE takes place for a negligible fraction of MHs, which
is consistent with the current observations: no EMP stars have so far been observed.


\item faint supernova (FSN) models, that reproduce the abundance ratio of CE-EMP stars, 
can also successfully account for the absolute iron/carbon abundance of those stars 
in the recollapsing regions.

\item we can therefore conclude that
the enrichment mode IE of MHs from a single CCSN or FSN event can explain the formation of CN- and CE-EMP stars.

\end{itemize}

These findings are compared with the observations of EMP stars in the halo region 
of our Galaxy, and ultra-faint dwarfs (UFDs).
Although our knowledge is now limited to a single event of transition from Pop III star formation to the succeeding generation of stars, i.e., the very first step of the matter cycle in ISM, 
we will extend this study to the early-generations of galaxies in which a number of cycles take place and 
directly observed by the next-decade instruments such as {\it JWST} and TMT.

\section*{acknowledgments}

We are grateful for the anonymous referee to the suggestions and comments
for our work.
We also thank T. Nozawa and M. Ishigaki who kindly give us the SN models.
Thanks to the deep insight of R. Schneider, L. Graziani, N. Tominaga, and T. Ishiyama, we can improve our manuscript.
GC and SH are supported by Research Fellowships of the Japan
Society for the Promotion of Science (JSPS) for Young Scientists.
HS acknowledges the financial supports from
JSPS Grant-in-Aid for Scientific Research (17H01101,17H02869,17H06360).
The numerical simulations in this work are carried out 
on Cray XC30 at Center for Computational Astrophysics, National Astronomical Observatory of Japan and
on Cray XC40 at the Yukawa Institute Computer Facility.




\end{document}

%% file: allvars.tex
\def\vb#1{\mbox{\boldmath $#1$}}

\newcommand{\Gadget}{{\sc gadget}}

\newcommand{\percc}{{\rm cm^{-3}}}

\newcommand{\E}[1]{\times 10^{#1}}

\newcommand{\nH}{n_{{\rm H}}}
\newcommand{\mH}{m_{{\rm H}}}
\newcommand{\nHcen}{n_{\rm H, cen}}
\newcommand{\Zcen}{Z_{\rm cen}}
\newcommand{\Zhalo}{Z_{\rm halo}}

\newcommand{\HII}{H~{\sc ii}}
\newcommand{\zcol}{z_{\rm col}}
\newcommand{\Mhalo}{M_{\rm halo}}
\newcommand{\Rhalo}{R_{\rm halo}}
\newcommand{\Ebin}{E_{\rm bin}}
\newcommand{\Mjeans}{M_{\rm J}}

\newcommand{\MPopIII}{M_{\rm Pop III}}
\newcommand{\Esn}{E_{{\rm SN}}}
\newcommand{\erg}{\rm erg}

\newcommand{\tlife}{t_{\rm life}}
\newcommand{\trecol}{t_{\rm recol}}

\newcommand{\Mmet}{M_{\rm met}}
\newcommand{\Nmet}{N_{\rm met}}

\newcommand{\QH}{Q({\rm H})}

\newcommand{\mpart}{m_{\rm p}}
\newcommand{\Mmin}{M_{\rm min}}
\newcommand{\Rcr}{{\cal R}_{\rm cr}}
\newcommand{\Nngb}{N_{\rm ngb}}

\newcommand{\Fe}{{\rm Fe}}

\newcommand{\abH}[1]{{\rm [{#1}/H]}}

\newcommand{\abFe}[1]{{\rm [{#1}/Fe]}}

\newcommand{\XH}{X_{{\rm H}}}
\newcommand{\Zsun}{{\rm Z_{\bigodot}}}
\newcommand{\Msun}{{\rm M_{\bigodot}}}
\newcommand{\Mcut}{M_{\rm cut}}

\defcitealias{Chiaki15}{C15}
\defcitealias{Hirano14}{H14}
\defcitealias{Ishigaki14}{I14}
\defcitealias{Tominaga14}{T14}
\defcitealias{Marassi15}{M15}